\documentclass[aps,prd,twocolumn,superscriptaddress,nofootinbib,longbibliography,floatfix]{revtex4-2}

\usepackage[utf8]{inputenc}
\usepackage[T1]{fontenc}
\usepackage{amsmath,amssymb,amsfonts,mathtools}
\usepackage{bm}
\usepackage{braket}
\usepackage{graphicx}
\usepackage{dsfont}
\usepackage{booktabs}
\usepackage{xcolor}
\usepackage[colorlinks=true,linkcolor=blue,citecolor=blue,urlcolor=blue]{hyperref}
\newcommand{\half}{\frac{1}{2}}
\newcommand{\R}{\mathbb{R}}
\newcommand{\Z}{\mathbb{Z}}
\newcommand{\SLR}{\mathrm{SL}(2,\R)}
\newcommand{\SLp}{\mathrm{SL}^{+}(2,\R)}
\newcommand{\slR}{\mathfrak{sl}(2,\R)}
\newcommand{\Ad}{\mathrm{Ad}}
\newcommand{\Pk}[1]{\mathcal{P}_{#1}}        
\newcommand{\Dm}[1]{\mathcal{D}^{-}_{#1}}    
\newcommand{\Cas}{\mathcal{C}}

\newcommand{\thop}{t_{\rm h}}                
\newcommand{\Hbare}{H_{\rm bare}}
\newcommand{\Hphys}{H_{\rm phys}}
\newcommand{\Nmax}{n_{\rm max}}

\newcommand{\dd}{\mathrm{d}}

\newcommand{\id}{\mathds{1}}

\begin{document}

\title{Non-compact Kogut-Susskind gauge theories: 1D lattice models with exact $\SLR$ gauge invariance}

\author{Carlos Fulgado-Claudio}
\email{carlos.f.claudio99@gmail.com}
\affiliation{Instituto de F\'isica Te\'orica, UAM-CSIC, Universidad Autónoma de Madrid, Cantoblanco, 28049 Madrid, Spain}

\author{Torsten V. Zache}
\affiliation{Institute for Theoretical Physics, University of Innsbruck, Innsbruck, 6020, Austria}
\affiliation{Institute for Quantum Optics and Quantum Information of the Austrian Academy of Sciences, Innsbruck, 6020, Austria}

\date{\today}

\begin{abstract}
The Kogut-Susskind (KS) formulation of gauge theories constructs quantum many-body Hamiltonians with gauge invariance under a compact Lie group exactly implemented on a spatial lattice.
Here, we revisit this construction for the case of a \emph{non-compact} gauge group.
Although motivated in part by theories like general relativity whose first-order formulation involves the non-compact Lorentz group as a local symmetry, we focus on much simpler cases in one spatial dimension akin to the  Schwinger model.
In a bottom-up approach, we show how exact gauge invariance under the simplest non-compact Lie group $\SLR$ can be retained exactly on a 1D lattice, and highlight the differences that arise with respect to compact gauge groups, including a continuous spectrum of the involved Casimir operators of the gauge fields and the emergence of infinite-dimensional representations of the matter fields. 
For the pure gauge theory, we discuss how a consistent restriction to the positive semigroup $\SLp$ reproduces well-known results from the quantization of Jackiw--Teitelboim (JT) gravity.
Motivated by the constraint structure of general dilaton gravity, we introduce static charges coupled to the non-compact gauge field, which can be realised by single-fermion degrees of freedom living in an additional, synthetic dimension.
With static charges our lattice model remains exactly integrable, but when extending to dynamical fermionic matter, we arrive at an interacting 1D $\SLR$ KS gauge theory.
We give an interpretation of this model as a chain of charges sequentially connected by strings whose tension pulls matter left or right depending on the sign of the relative continuous $\SLR$ fluxes.
This picture becomes exact when the involved $\SLR$ charges are strongly self-interacting, where we solve the reduced two-body problem analytically.
Our work thus lays the basis for future studies of interacting non-compact gauge theories using tools that benefit from a Hamiltonian lattice formulation such as classical tensor network techniques or quantum simulations.  
\end{abstract}

\maketitle

\section{Introduction}
\label{sec:intro}

The Kogut-Susskind (KS) Hamiltonian formulation~\cite{Kogut:1974ag} has become the prevalent language for the quantum simulation of gauge theories~\cite{Wiese:2013uua,Banuls:2019bmf,Bauer:2022hpo,DiMeglio:2023nsa,Halimeh:2023lid}.
Over the last decades, many models with Abelian gauge groups have been experimentally realised with systems based on trapped ions~\cite{Martinez:2016yna,Nguyen:2021hyk,Meth:2023wzd,De:2024smi}, cold atoms in optical lattices~\cite{Schweizer:2019lng,Gorg:2019dxi,Yang:2020yer,Zhou:2021kdl}, atomic gases~\cite{Mil:2019pbt,Frolian:2022chiral}, Rydberg arrays~\cite{Bernien:2017ubn,Semeghini:2021wls,Gonzalez-Cuadra:2024xul} (see~\cite{Surace:2019dtp} for the LGT interpretation of \cite{Bernien:2017ubn}), and superconducting qubits~\cite{Klco:2018kyo,Mildenberger:2022jqr,Farrell:2023fgd,Cochran:2024rwe,Cobos:2025krn}.
The challenge to bring also gauge theories based on non-Abelian groups within experimental reach has created a renewed interest in the KS formulation itself and a significant range of modifications and variants has been addressed theoretically~\cite{Byrnes:2005qx,Banerjee:2012xg,Zohar:2012xf,Tagliacozzo:2012df,Zohar:2014qma,Raychowdhury:2018osk,Davoudi:2020yln,Buser:2020cvn,Ciavarella:2021nmj,Gonzalez-Cuadra:2022hxt,Zache:2023dko,Hayata:2023bgh,Halimeh:2024bth,Alexandru:2023qzd,Siew:2025thj}, with first implementations on quantum hardware~\cite{Klco:2019evd,Atas:2021ext,Farrell:2022wyt,Farrell:2022vyh,Ciavarella:2024fzw,ARahman:2021ktn,John:2026gut,Ilcic:2026cac,Kaji:2026znd}.
A common mathematical feature of most approaches is the use of compact groups, which simplifies the involved representation theory in such a way that the resulting gauge field fluxes are naturally discrete.
In this article, we explore a generalization of the KS formulation with non-compact gauge groups.

Non-compact gauge groups arise naturally in physical models of gravity.
General relativity, when expressed in its first-order formulation, can also be understood as a gauge theory involving the non-compact Lorentz group~\cite{Utiyama:1956sy,Kibble:1961ba}.
This fact is the basis for several ongoing constructions targeting a theory of quantum gravity: Euclidean lattice gravity was formulated based on Poincaré and de Sitter groups in the path-integral language~\cite{Smolin:1978yb,Das:1978ni,Menotti:1986vg,Menotti:1987ka,Caselle:1987yk}; spin networks and spin foams for Lorentzian gravity are built on the principal series of ${\rm SL}(2,\mathbb{C})$~\cite{Freidel:2002xb,Perez:2000ec,Livine:2002ak,Engle:2007wy,Barrett:2009mw}.
Non-compactness also persists in lower dimensional models of gravity, e.g. in the SU(1,1) spin networks of 2+1 dimensional Lorentzian gravity~\cite{Freidel:2002hx}, where the reduced number of gravitational degrees of freedom simplifies the physics.
We emphasize that we do not attempt to construct models of quantum gravity in 3+1D (or 2+1D) here.
Instead we are interested in generalizations of conventional KS lattice gauge theories based on a non-compact gauge group, which appears to be largely unexplored.
On a technical level, one can anticipate constrained quantum many-body models with matter and gauge field degrees of freedom carrying appropriate unitary infinite-dimensional representations of the non-compact gauge group.
To avoid confusion, we note that the term ``non-compact'' has also been used for non-compact lattice formulations and continuous-variable encodings of compact gauge groups~\cite{Diekmann:1993rm,Ale:2024uxf,Mendicelli:2026ncv}.

The anchor for our exploration is Jackiw--Teitelboim (JT) gravity~\cite{Jackiw:1984je,Teitelboim:1983ux}, which involves a metric in two space-time dimensions coupled to a scalar field, the so-called dilaton, which describes the near-horizon dynamics of near-extremal black holes~\cite{Almheiri:2014cka,Maldacena:2016upp,Mertens:2022irh}.
In its first-order formulation, the classical field theory matches a BF gauge theory with algebra $\mathfrak{sl}(2,\mathbb{R})$~\cite{Fukuyama:1985gg,Isler:1989hq,Chamseddine:1989yz}, which is a topological model with no propagating degrees of freedom in the bulk such that its dynamics is fully captured at the space-time boundary.
Throughout this article, we will heavily rely on group-theoretic methods that have been used in the quantization of JT gravity to derive closed expressions for many observables~\cite{Blommaert:2018oro,Blommaert:2018iqz,Iliesiu:2019xuh,Mertens:2017mtv,Kitaev:2018wpr,Yang:2018gdb,Mertens:2022irh}.
Recently, the Hilbert space of JT gravity with matter has been characterised in terms of the fluxes between matter insertions~\cite{Lin:2019qwu,Lin:2022rbf,Harlow:2021dfp,Penington:2023dql,Kolchmeyer:2023gqz,Iliesiu:2024cnh}, a structure that strongly suggests a lattice description.
A discretisation of the BF formulation of JT has also been studied, on hyperbolic tilings of Euclidean AdS2 where it maps to an Ising model~\cite{Erdmenger:2024jsb}, or at the level of holonomies and boundary symmetries~\cite{Ozer:2026jt}.
Guided by these established results, we devise a bottom-up KS construction which to the best of our knowledge has not been attempted previously.

Another inspiration for this work is the hope to quantum simulate gravity in the lab, which has many intriguing perspectives~\cite{Brown:2019hmk,Nezami:2021yaq}.
Recent proposals to quantum simulate JT gravity employ its holographic boundary description, which also arises as the low-energy effective theory of the Sachdev--Ye--Kitaev (SYK) model~\cite{Garcia-Alvarez:2016wem,Danshita:2016xbo,Pikulin:2017mhj,Babbush:2018mia,Luo:2017bno,Jafferis:2022crx,Kobrin:2023rfr,Shapoval:2022xeo,Asaduzzaman:2023wtd}.
A KS formulation of JT gravity provides a complementary angle to this idea, which is conceptually closer to the way the lattice Schwinger model realizes QED2 directly as an ordinary $1+1$ dimensional lattice model.
Our constructions could thus also form the basis for an alternative approach toward quantum simulations of JT gravity via its lattice gauge theory formulation, where exact results from the integrable continuum parent can be used as benchmarks.

In the remainder of this introduction, we give an overview of the structure of this article, summarizing our results and indicating where we differ from an established gravitational interpretation.

First, in Sec.~\ref{sec:ks} we review the Kogut--Susskind construction for compact groups, and  identify which specific elements change for the non-compact ${\rm SL}(2,\mathbb{R})$ case. In particular, we set up the pure gauge theory by diagonalising an elliptic, compact generator of ${\rm SL}(2, \mathbb{R})$, such that unitary representations on the lattice links give an infinite ladder of gauge field states on which the generators act as a ``nearest-neighbour'' hopping.
Restricting to the positive semigroup and adopting the boundary gravitational condition of Ref.~\cite{Blommaert:2018iqz}, this approach reproduces the JT partition function. 

Then, in Sec.~\ref{sec:matter}A, we introduce matter fields as static charges on the lattice sites, motivated by the constraint structure of two-dimensional dilaton gravity.
We choose the static charges to live in the discrete series of ${\rm SL}(2, \mathbb{R})$, such that they can be realised by fermionic degrees of freedom living in a synthetic dimension.
In direct analogy to the well-known elimination of gauge fields in other 1D gauge theories, such as the Schwinger model, dressing the charges and fluxes with the gauge field holonomy effectively eliminates the link variables.
This shows that the theory 
can be completely solved by representation theory, with every region between two charges carrying a conserved principal series label, and Gauss's law gluing these regions.
At this point, gauge invariance alone allows for different a priori inequivalent electric Hamiltonians, and we define a family of models that interpolate between contributions that live only at the boundary (like in JT gravity) or homogeneously on the whole lattice (as in a conventional KS formulation). 
Notably, the constraints among charges and gauge fields imposed by Gauss' law lead to a distribution of boundary links given by a continuous dual Hahn polynomial weight that contains the JT density of states and the matter vertex. 

Finally, in Sec.~\ref{sec:matter}B, we introduce further gauge-invariant terms in the lattice Hamiltonian that allow the matter charges to move, or to self-interact.
At strong self-interaction, double occupancies are energetically forbidden, such that gauge fields are uniquely specified by region labels that remain conserved despite their precise spatial positions being dynamical.
As a result the model turns into a chain of dynamically coupled Wannier--Stark ladders whose force is given by the differences of neighbouring Casimir labels.
A single charge then performs Bloch oscillations, and in the case where the charge has a well-defined internal state, these oscillations dephase due to the frequencies of oscillation being averaged according to the JT vertex.
For any neighboring pair of charges, the interior label gives rise to a string tension between them, which, depending on the tension of the boundary regions, leads to an effective attraction or repulsion.
Without double occupation, the two-charge problem is exactly solvable, and determines when the charges oscillate independently and when they form a bound pair; with double occupation, the interior labels become dynamical.

The article concludes with a summary and outlook on future directions in Sec.~\ref{sec:conclusions}.
In order to remain self-contained, we collect
facts about the unitary representations of ${\rm SL}(2,\mathbb{R})$, the map between the first-order formulation of JT and a BF gauge theory, and details on the scaling of different quantities with our numerical truncations in the Appendices.

\section{Pure $\SLR$ lattice gauge theory}
\label{sec:ks}

In this section we construct a 1D Hamiltonian lattice gauge theory (LGT) whose gauge group is $\SLR$, the non-compact group of real $2\times2$ matrices of unit determinant.
We first recall important elements of the Kogut--Susskind construction for a general gauge group (Sec.~\ref{sec:compact}).
Then we fix our conventions for $\SLR$ (Sec.~\ref{sec:algebra}), identify the basis in which the link maps naturally to an LGT (Sec.~\ref{sec:bases}), and build the link Hilbert space and the physical states of the pure gauge theory (Secs.~\ref{sec:links} and~\ref{sec:pure}); the unitary representations are reviewed in Appendix~\ref{sec:irreps}. 
Finally, we show that the pure gauge theory reproduces the JT partition function (Sec.~\ref{sec:jt}).

\subsection{The compact case: Kogut--Susskind construction}
\label{sec:compact}

We consider a 1D lattice, i.e. a chain of $N_{\rm links}$ links with $N_{\rm links}-1$ sites/vertices between them, labelled as shown in Fig.~\ref{fig:1}{\bf(a)}.
Each link carries gauge fields parametrized by group elements $g$, meaning its Hilbert space $L^2(G)$ is the space of square-integrable functions (with respect to the Haar measure) $\psi(g)$, where $G$ is the gauge group of the LGT.
This local Hilbert space can be spanned by the so-called group basis, $\ket{g}$, such that $\psi(g)=\braket{g|\psi}$.
We can also work in a dual basis: for a specific irreducible representation $\lambda$, the states $\ket{\lambda,m}$ form a vector space $V_\lambda$. Group elements are represented on this vector space as $D^\lambda(g)$.
This defines operators via the matrix elements $\braket{\lambda,m|D^\lambda(g)|\lambda,n}=D^\lambda_{m,n}(g)$ which, regarded as functions of $g$, form a dual basis for $L^2(G)$ which we denote as $\ket{\lambda, m, n}$.
We will refer to the resulting isomorphism $L^2(G) \simeq \bigoplus_{\lambda} V_\lambda \otimes V^*_\lambda$ as the Peter-Weyl theorem.

In order to construct the Hamiltonian of the LGT, we first introduce the operators that generate group multiplication of states in the group basis. Specifically, left and right multiplication is implemented by operators $\Theta^\alpha_h$, with $\alpha=L, R$, acting as $\Theta^L_h\ket{g}=\ket{h^{-1}g}$ and $\Theta^R_h\ket{g}=\ket{gh}$. Writing the group element $h$ as $h=e^{-i\theta^a K_a}$, these operators are expressed as $\Theta^L_h=e^{i\theta^aL_a}$ and $\Theta^R_h=e^{i\theta^aR_a}$. We refer to $L_a$ and $R_a$ as \emph{left} and \emph{right electric field}. Their action on either basis is  given by

\begin{equation}
\begin{gathered}
    L_a\ket{g}=i\frac{d}{dt}\ket{e^{-itK_a}g}\Bigg|_{t=0},\\
    R_a\ket{g}=-i\frac{d}{dt}\ket{ge^{-itK_a}}\Bigg|_{t=0},\\
    L_a\ket{\lambda,m,n}=-\sum_p\big(K^{(\lambda)}_a\big)_{mp}\ket{\lambda,p,n},\\
R_a\ket{\lambda,m,n}=\sum_p\big(K^{(\lambda)}_a\big)_{pn}\ket{\lambda,m,p},\end{gathered}
\label{eq:LR-action}
\end{equation}
It is important to note that $L_a$ and $R_a$ are not independent, and for any group their relationship follows from $R_a\ket{g}=-(\Ad_g)^b{}_aL_b\ket{g}$, where the adjoint matrix is defined by $ gJ_ag^{-1}=(\Ad_g)^b{}_aJ_b$. This fact will be fundamental when imposing Gauss's law further below. 

In one spatial dimension, gauge fields have no propagating degrees of freedom due to the absence of ``magnetic'' plaquette interactions. 
The pure-gauge Hamiltonian therefore contains only ``electric'' energy terms, given by the quadratic Casimir operators $\Cas\equiv\Cas(L)=\Cas(R)$.
Crucially, any local Casimir is gauge-invariant, $\left[\Cas_x, G_{y,a}\right]$ =0, where $G_{x,a} = L_{x+1,a} + R_{x,a}$ denote the Gauss-law operators.
One can thus construct a gauge-invariant Hamiltonian by summing the Casimir of either left- or right-electric field on each link,
\begin{equation}
H_{\rm bare}=\sum_{\rm links}\Cas(L),
\label{eq:Hbare}
\end{equation}
up to normalization.
We note, however, that alternative Hamiltonians can be defined if one does not insist on translation invariance.
By summing $\Cas(L)$ over only a subset of the links, we will later see that the presence of matter can lead to different results. 
This will become particularly relevant when connecting to JT gravity.

\subsection{The non-compact case: $\SLR$}
\label{sec:noncompact}

\subsubsection{Algebra and conventions}
\label{sec:algebra}

The Lie algebra $\slR$ is spanned by three generators that satisfy
\begin{equation}
  [J_0,J_1]=J_2,\qquad [J_0,J_2]=J_1,\qquad [J_1,J_2]=J_0.
  \label{eq:real-algebra}
\end{equation}
A specific representation of $\slR$ is the set of real matrices $J_0=\half\sigma^z$, $J_1=\frac{i}{2}\sigma^y$ and $J_2=\half\sigma^x$. Note that, being finite-dimensional, this representation is non-unitary. The invariant Killing form is given by $\eta_{ab}=2\,\mathrm{tr}(J_aJ_b)=\mathrm{diag}(1,-1,1)$, which is used to classify a real combination $X=x^aJ_a$ as elliptic, hyperbolic or parabolic depending on whether $\eta_{ab}x^ax^b$ is negative, positive or zero, respectively. 
Thus $J_1$ is classified as elliptic and $J_0$ and $J_2$ are hyperbolic. While the former generates the compact subgroup of rotations, the latter two both generate non-compact subgroups.

In a unitary representation $U(e^{\theta^aJ_a})=e^{\theta^a\,\mathrm{d}U(J_a)}$of the group, the generators $\mathrm{d}U(J_a)$ are anti-Hermitian elements of the algebra.
Throughout this work, we will instead employ Hermitian operators $K_a=iJ_a$.
Diagonalising any of them defines a basis of the corresponding representation space. 
In contrast to the compact case, for a non-compact group this choice has direct implications which will be discussed in the following Sec.~\ref{sec:bases}.
In this work, we mainly stick to the \emph{elliptic basis}, where $K_1$ is diagonal. 
The combinations $K^\pm=K_0\mp iK_2$ of the hyperbolic generators then act as ladder operators,
\begin{equation}
  [K_1,K^\pm]=\pm K^\pm,\qquad [K^+,K^-]=-2K_1 ,
  \label{eq:K-algebra}
\end{equation}
and the quadratic Casimir operator reads
\begin{equation}
  \Cas=\eta^{ab}J_aJ_b=K_1^2-\tfrac12\{K^+,K^-\}.
  \label{eq:casimir}
\end{equation}
By definition, the Casimir commutes with all generators, and its eigenvalues classify the irreducible representations.
We parametrise these eigenvalues as $\lambda(\lambda-1)$.
Within a representation the states are labelled by the eigenvalue of $K_1$,
\begin{align}
\Cas\ket{\lambda,m}=\lambda(\lambda-1)\ket{\lambda,m}, &&
  K_1\ket{\lambda,m}=m\ket{\lambda,m}.
\label{eq:basis}
\end{align}
Because $e^{4\pi J_1}=1$ in $\SLR$, the weights obey $2m\in\Z$, that is $m\in\mu+\Z$ with $\mu\in\{0,\half\}$. Throughout this work we choose $\mu=0$.

\subsubsection{Hyperbolic versus elliptic basis}
\label{sec:bases}

For our purposes, the choice of the diagonal generator has two crucial consequences: {\it (i)} the spectrum of non-compact generators is continuous, while the spectrum of the compact $K_1$ is discrete; and {\it (ii)} the ladder operators constructed from the two remaining generators act differently on the states due to the following 
elementary fact.
If $[H,A]=\alpha A$, then $A$ maps an eigenvector of $H$ with eigenvalue $h$ to one with eigenvalue $h+\alpha$.
Suppose we diagonalise a hyperbolic generator, $K_0\ket{s}=s\ket{s}$.
Then $K_0$ generates a non-compact subgroup, so no periodicity condition quantises its eigenvalue, such that the spectrum is the real line, $s\in\R$, and the ladder operators 
$\tilde K^\pm=K_1\pm K_2$, which obey
\begin{equation}
  [K_0,\tilde K^\pm]=\pm i\,\tilde K^\pm ,
  \label{eq:imag-shift}
\end{equation}
change $s$ by an imaginary unit.
The matrix elements are 
thus 
of the schematic form
\begin{equation}
    \braket{s'|\tilde K^{\pm}|s}\sim F_{\pm}(s)\delta(s'-s\mp i) 
    \;.
\end{equation}
While these are well-defined albeit unbounded operators, they have vanishing matrix elements between the points of a real grid.
Therefore a discretisation of $s$ does not carry the algebra even approximately.

In the elliptic basis both obstacles are absent. The spectrum of $K_1$ is discrete, the shifts are $\pm1$ as per Eq.~\eqref{eq:K-algebra}, and by Eq.~\eqref{eq:principal} the generators $K_0=\half(K^++K^-)$ and $K_2=\frac{i}{2}(K^+-K^-)$ are tridiagonal.
The weight $m$ is then a site index along an infinite ladder, on which the generators act as nearest-neighbour ``hopping'', and the representation can be readily truncated by restricting $m$ to a finite interval (see Sec.~\ref{sec:schemes}).
For a compact group this ladder is finite, while here it is infinite, which is the remaining price to pay for non-compactness.

\subsubsection{Link Hilbert space for $\SLR$}
\label{sec:links}

For a non-compact group, the Peter--Weyl theorem of compact group LGTs is replaced by a generalization called the Plancherel theorem~\cite{HarishChandra:1952,Lang:1985}.
Schematically, $L^2(G) \simeq \int^\oplus_{\lambda} V_\lambda \otimes V^*_\lambda$, where the direct sum over irreps in general contains contributions from a direct integral over the unitary irreps that are weighted by the so-called Plancherel measure.
For the compact example of $\mathrm{SU}(2)$, the irreps $j$ of dimension $2j+1$ all appear with normalized basis states $\ket{j,m,m'}$ with an equal weight.
For the non-compact case of $\SLR$ the decomposition contains all irreps of the principal series, with density $\rho(k)=k\tanh(\pi k)$ for $\mu=0$ and $\rho(k)=k\coth(\pi k)$ for $\mu=\tfrac12$, as well as irreps in the discrete series $\mathcal D^\pm_\ell$, with weights proportional to $2\ell-1$ (see App~\ref{sec:irreps} for details, and Eq.~\eqref{eq:plancherel-SLR} further below).

In principle, the gauge field degrees of freedom that we place on the links of our lattice could live either in the principal series, with $\Cas(L)=-(k^2+1/4)$, or in one of the discrete series $\mathcal{D}^\pm_\ell$, with $\Cas(\ell)=\ell(\ell-1)$.
However, we will only consider the case where link degrees of freedom live in the principal series, so $\Cas(L)\ket{k,m,n}=-(k^2+\frac14)\ket{k,m,n}$, which will give rise to a fully consistent construction.  
This choice can be motivated by results obtained in the context of JT gravity~\cite{Blommaert:2018iqz}, and can be interpreted as a restriction to the subsemigroup $\SLp$, whose Plancherel decomposition does not include the discrete series.
In order to make the electric energy spectrum positive and bounded from below by $0$, we swap an overall sign w.r.t. the ordinary KS construction and define a new \emph{physical} Hamiltonian in terms of $H_{\rm bare}$ of Eq.~\eqref{eq:Hbare} as
\begin{equation}
    H_{\rm phys}=-\frac{H_{\rm bare}}{N_{\rm links}}-\frac14.
    \label{eq:Hphys}
\end{equation}
We will come back to the meaning of this choice, including the scaling factor $N_{\rm links}$, in Sec.~\ref{sec:H-choice} below.

\begin{figure*}
    \centering
    \includegraphics[width=0.95\linewidth]{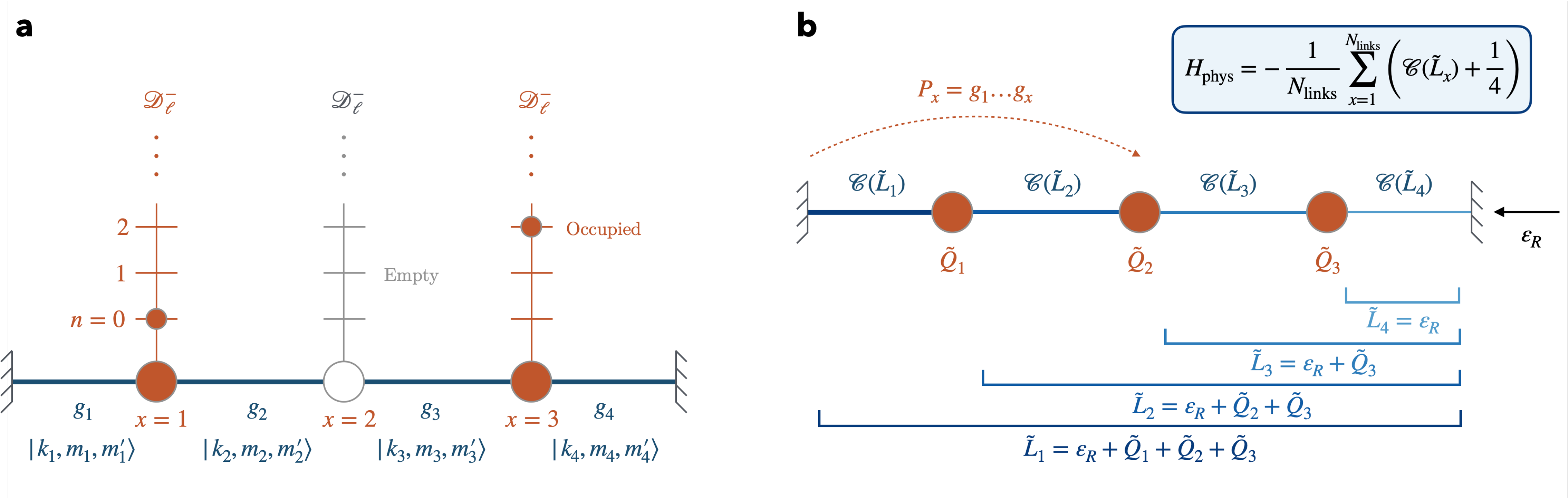}
    \caption{{\bf (a)} Chain of $N_{\rm links}$ links carrying $\ket{k,m,m'}$ and $N_{\rm links}-1$ vertices, each carrying a $\Dm{\ell}$ tower $n=0,1,\dots$ (synthetic dimension) that is empty or occupied. {\bf (b)} Schematic of the Hamiltonian. One Casimir per link's flux, with the fluxes $\tilde{L}_j$ constructed with accumulated charges from the right end, and the dressing of the charge at $x$ is constructed from the holonomy $P_{x}$.}
    \label{fig:1}
\end{figure*}

\subsubsection{Pure gauge theory: physical states}
\label{sec:pure}

For the pure gauge theory, i.e. in the absence of matter, the Gauss-law operator $G_{x,a}=R_{x,a}+L_{x+1,a}$ at vertex $x$ is imposed 
on physical states, $G_{x,a}\ket{\psi_{\rm phys}}=0$ for all $x$ and $a$.
To illustrate the consequences of this constraint, consider first two links meeting at a vertex with state $\ket{\psi}=\ket{k_1,m_1,m'_1}\ket{k_2,m_2,m'_2}$. Since $\Cas(L_x)=\Cas(R_x)=-(k_x^2+\frac14)$, and using the fact that operators on different links commute, physical states satisfy the identity $\Cas(R_x)\ket{\psi_{\rm phys}}=-\eta^{ab}R_{x,a}R_{x,b}\ket{\psi_{\rm phys}}=\eta^{ab}R_{x,a}L_{x+1,b}\ket{\psi_{\rm phys}}=\eta^{ab}L_{x+1,b}R_{x,a}\ket{\psi_{\rm phys}}=-\eta^{ab}L_{x+1,b}L_{x+1,a}\ket{\psi_{\rm phys}}=\Cas(L_{x+1})\ket{\psi_{\rm phys}}$, forcing $k_x=k_{x+1}$.
On the other hand, the elliptic component of $G$ has eigenvalue $m'_1-m_2$.
Therefore 
physical states must satisfy $m'_1=m_2$. 
The remaining 
constraints require the coefficients of all values of the contracted index to be equal. 
Iterating over the vertices of the chain, we see that all physical states take the form
\begin{equation}
\begin{aligned}
  \ket{\psi_{\rm phys}}&=\int\!\dd k\,\rho(k)\sum_{m,m'}\phi(k,m,m')\\
  &\times\sum_{\{\sigma\}}\ket{k,m,\sigma_1}\ket{k,\sigma_1,\sigma_2}\cdots\ket{k,\sigma_{N_{\rm links}-1},m'},
\end{aligned}
  \label{eq:phys-states}
\end{equation}
where $\rho(k)$ is the Plancherel measure. In that sense, only the Casimir label $k$ common to all links and the two boundary indices $m$ and $m'$ remain in the wavefunction $\phi(k,m,m')$.
A complete basis for the physical Hilbert space is given by $\ket{k,m,m'}\rangle\equiv\sum_{\{\sigma\}}\ket{k,m,\sigma_1}\ket{k,\sigma_1,\sigma_2}\cdots\ket{k,\sigma_{N_{\rm links}-1},m'}$, with $H_{\rm phys}\ket{k,m,m'}\rangle=k^2\ket{k,m,m'}\rangle$, where we write $\ket{k,m,m'}\rangle$ for the state of the whole lattice to distinguish it from the states $\ket{k,m,m'}$ of individual links. 
Note that the projection of this state onto the group basis leads to the composition $\sum_\sigma D^k_{m\sigma}(g_1)D^k_{\sigma n}(g_2)=D^k_{mn}(g_1g_2)$.
This means that physical states are a function of the global holonomy $W=g_1g_2\cdots g_{N_{\rm links}}$ alone, with no local degrees of freedom left.
This structure directly parallels the case of compact groups, except for the continuity of the electric ``flux'' labels $k$.

\subsection{Equivalence with Jackiw--Teitelboim quantum gravity}
\label{sec:jt}
We have seen that the pure lattice gauge theory reduces to a single quantum mechanical degree of freedom.
We now discuss its connection to the Jackiw--Teitelboim theory. 
This makes use of the equivalence between JT gravity and an $\slR$ BF theory, which is reviewed in Appendix~\ref{sec:JT}.
Here we focus on the boundary dynamics and show that the pure gauge theory recovers the exact JT disk partition function (Sec.~\ref{sec:bench-Z}). The connection remains in the presence of matter, as we will discuss further below in Sec.~\ref{sec:bench-fusion}.

In the metric formulation of JT gravity (Appendix~\ref{sec:JT}) 
the bulk Hamiltonian vanishes on the constraint surface, such that all the dynamics of the theory resides at the boundary. 
A well-posed variational problem requires the Gibbons--Hawking--York term,
\begin{equation}
  S_\partial=2\int_{\partial\mathcal M}\!\dd t\,\sqrt{-\gamma}\,\phi\,(K-1),
\end{equation}
where $\gamma$ is the induced metric on the boundary and $K$ its extrinsic curvature.
The equation of motion of $\phi$ sets $R=-2$, so only $S_\partial$ survives on shell. 
Imposing, close to the AdS boundary, $\sqrt{-\gamma}=1/\epsilon$ and $\phi|_{\partial\mathcal M}=\phi_b/\epsilon$, where $\phi_b$ is the renormalised boundary value of the dilaton, and parametrising the boundary curve by its Poincaré time $F(t)$, $S_\partial$ reduces at leading order in $\epsilon$ to the so-called Schwarzian action~\cite{Maldacena:2016upp,Mertens:2022irh}
\begin{equation}
  S_\partial=-2\phi_b\int\dd t\,\{F,t\},\qquad
  \{F,t\}=\frac{\dddot{F}}{\dot{F}}-\frac32\Big(\frac{\ddot{F}}{\dot{F}}\Big)^2 ,
\end{equation}
where $\{F,t\}$ is the Schwarzian derivative and dots represent time derivatives.

Remarkably, JT gravity has been successfully quantized using path integral methods~\cite{Stanford:2017thb,Saad:2019lba}.
Here, we only quote the final result for $\mathcal{M}$ a Euclidean disk, where the partition function can be exactly evaluated in terms of the Schwarzian as
\begin{equation}
\label{eq:Zdisk}
\begin{aligned}
Z_{\rm disk}(\beta)&\equiv\int\frac{\mathcal{D}g\,\mathcal{D}\phi}{\rm Vol}\, e^{-I_{\rm JT}}
=\int\frac{\mathcal{D}F}{{\rm SL}(2,\mathbb{R})}\, e^{-I_\partial}\\
&\propto\int_0^\infty dk\,k\sinh(2\pi k)\,e^{-\beta k^2/(4\phi_b)}
\\&=4\pi^{3/2}\left(\frac{\phi_b}{\beta}\right)^{3/2}e^{4\pi^2\phi_b/\beta}.
\end{aligned}
\end{equation}
Our bottom-up KS construction was designed to exactly reproduce this result, as we discuss next.

\subsubsection{The positive semigroup and its Plancherel measure}
\label{sec:semigroup}
A fundamental insight to match a Hamiltonian lattice gauge theory to JT gravity 
comes from Ref.~\cite{Blommaert:2018iqz}, where the matrix group of $\SLR$ was restricted to $\SLp$ by only considering matrices with positive entries.
The result is closed under multiplication but not under inversion, so it forms a subsemigroup and not a subgroup.

The decomposition of square-integrable functions into unitary irreps, whose density is the Plancherel measure (cf. Sec.~\ref{sec:links}), changes in two ways when going from $\SLR$ to $\SLp$. For integer weights ($\mu=0$) and up to normalisation~\cite{Blommaert:2018iqz},
\begin{align}
  L^2\big(\SLR\big)&\simeq\int_0^{\infty\,\oplus}\!\!\dd k\;k\tanh(\pi k)\;\mathcal{P}_k\otimes\mathcal{P}_k\nonumber\\
  &\quad\oplus\bigoplus_{\ell\geq1}(2\ell-1)\;\mathcal{D}^\pm_\ell\otimes\mathcal{D}^\mp_\ell ,
  \label{eq:plancherel-SLR}\\
  L^2\big(\SLp\big)&\simeq\int_0^{\infty\,\oplus}\!\!\dd k\;k\sinh(2\pi k)\;\mathcal{P}_k\otimes\mathcal{P}_k .
  \label{eq:plancherel_slp}
\end{align}

First, as already anticipated above, the discrete series is absent: functions on the semigroup are built from the principal series alone. 
Second, the principal series is weighted with $\rho(k)=k\sinh(2\pi k)$ instead of $k\tanh(\pi k)$: $k\tanh(\pi k)$ grows linearly, whereas $\rho(k)\simeq\frac12k\,e^{2\pi k}$ grows exponentially. In the context of JT, this exponential growth is connected to the Bekenstein--Hawking entropy of black holes~\cite{Mertens:2022irh}.
Since a link in $\mathcal{P}_k$ carries energy $k^2$ (Sec.~\ref{sec:links}), $\rho(k)$ is the correct density of states of JT gravity, which will allow us 
to match the JT partition function.

We emphasize that the absence of the discrete series concerns the gauge field, i.e. the wave-functions on the links.
It does not prevent matter from carrying discrete-series representations.
In fact, once again motivated by JT gravity, in which a matter field of conformal dimension $\ell$ enters as a Wilson line in $\mathcal{D}_\ell$~\cite{Blommaert:2018oro,Mertens:2022irh}, we will introduce charges living in the exact same irreps.

To implement the restriction to $\SLp$ in our construction of Hamiltonian lattice models, we adopt the following prescriptions. {\it(i)} The link Hilbert space contains only the principal series with $\mu=0$, weighted with $\rho(k)=k\sinh(2\pi k)$ wherever the Plancherel measure appears, e.g. in the resolution of the identity, in inner products and in traces. {\it(ii)} Since $\mathcal{P}_k\otimes\mathcal{D}_\ell^-$ contains elements within the discrete series, it may occur that, after solving Gauss's law, a certain link lives in a discrete irrep. When considering $\SLp$, we will exclude this situation either by considering only weights for which these states never appear in the first place (Sec.~\ref{sec:blocks}), or by explicitly projecting these states out.
Note that such a projector is a function of the link Casimirs, so it is gauge invariant and commutes with $H_{\rm phys}$. Consequently, this restriction is preserved in unitary dynamics and only concerns the choice of superselection sector of the initial state.

\subsubsection{Partition function}
\label{sec:bench-Z}
We finally test the equivalence between our lattice construction and continuum JT quantum gravity by computing the thermal partition function, $Z_\text{LGT} \sim \text{Tr} \left(e^ {-\beta H_\text{phys}}\right)$, of our LGT.
To evaluate the trace, we use the physical states of Sec.~\ref{sec:pure}, $|k,m,m'\rangle\rangle\equiv\sum_{\{\sigma\}}|k,m,\sigma_1\rangle|k,\sigma_1,\sigma_2\rangle\cdots|k,\sigma_{N_{\rm links}-1},m'\rangle$, whose wavefunction is $D^k_{mm'}(W)$.
Since they depend on the links only through the holonomy $W$, the calculation reduces to a single link, with the inner product 
\begin{equation}
    \langle\braket{k,m,m'|q,r,r'}\rangle_{\rm phys}=\frac{\delta(k-q)\,\delta_{mr}\,\delta_{m'r'}}{\rho(k)},
    \label{eq:phys-norm}
\end{equation}
where $\rho(k)$ is the Plancherel measure of the gauge group. 
Technically, this defines a reduced physical inner product, with the gauge-volume factors associated with the internal vertices removed: in the tensor product of the link Hilbert spaces, each of the $N_{\rm links}-1$ contracted indices contributes a divergent factor $\delta(k-k)\sum_\sigma\frac{1}{\rho(k)}$ to the norm, which is the volume $V$ of the gauge group at that vertex [see Eq.~\eqref{eq:regularisation} below]. By construction, the basis states diagonalise the Hamiltonian, $H_{\rm phys}\ket{k,m,m'}\rangle=k^2\ket{k,m,m'}\rangle$, for any $N_{\rm links}$.

The regularised partition function of the LGT is therefore defined as $Z_{\rm LGT}=V^{-1}\,\mathrm{Tr}(e^{-\beta H_{\rm phys}})$, where the factor $1/V$ is the quotient by the global group in the boundary path-integral prescription of Ref.~\cite{Blommaert:2018iqz}.
The trace over the free boundary indices $m$ and $m'$ contains divergent factors, which we regularise following Ref.~\cite{Blommaert:2018iqz},
\begin{equation}
    \begin{aligned}
        &\delta(k-k)=V_C, &&
        &\sum_m 1=\chi_k(\id)=\frac{V}{V_C}\rho(k),
    \end{aligned}
    \label{eq:regularisation}
\end{equation}
where $V$ is the regularised volume of the group and $V_C$ is the regularised volume of the space of conjugacy classes. The latter equation is a formal correspondence ${\rm dim} R\to\rho(k)$ between compact and non-compact groups~\cite{Blommaert:2018iqz,Mertens:2022irh}. Using Eq.~\eqref{eq:phys-norm}, we obtain
\begin{equation}
\begin{aligned}
    Z_{\rm LGT}&=\frac{1}{V}\int \dd k\,\rho(k)\\&\qquad\times\sum_{m,m'}\langle\braket{k,m,m'|e^{-\beta H_{\rm phys}}|k,m,m'}\rangle_{\rm phys}\\&=\frac{1}{V}\int \dd k\,\rho(k)e^{-\beta k^2}\sum_{m,m'}\frac{\delta(k-k)\delta_{m,m}\delta_{m',m'}}{\rho(k)}\\&=\frac{V}{V_C}\int \dd k\,\rho^2(k)e^{-\beta k^2}.
    \end{aligned}
\end{equation}

We are now fully back to the analogous continuum situation described previously in Ref.~\cite{Blommaert:2018iqz}.
There are two powers of $\rho(k)$ accounting for the two free elliptic indices at the boundaries of the lattice, $m$ and $m'$. The asymptotic AdS boundary condition imposes a parabolic constraint on one representation index, and in the reduced boundary-state normalisation of Ref.~\cite{Blommaert:2018iqz}, only the remaining index is traced, giving
\begin{equation}
    \begin{aligned}
    Z_{\rm LGT}^{\rm parabolic}&=\frac{V_C}{V}\int_0^\infty \dd k\,\Big(\sum_m 1\Big)\,e^{-\beta k^2}\\&=\int_0^\infty \dd k\,\rho(k)e^{-\beta k^2}.
    \end{aligned}
\end{equation}
Here, $\rho(k)$ is the Plancherel measure of the gauge group of the LGT. For the integer-weight principal-series sector of $\SLR$, the Plancherel density would be proportional to $k\tanh(\pi k)$, 
and we instead constrain to the subsemigroup $\SLp$ as outlined in Sec.~\ref{sec:semigroup}, which has Plancherel measure $\rho(k)=k\sinh(2\pi k)$, thus coinciding with the density of states of JT gravity.
We therefore finally arrive at
\begin{equation}
    Z_{\rm LGT}^{\rm parabolic}=\int_0^\infty \dd k\,k\sinh(2\pi k)e^{-\beta k^2}=\frac{\pi^{3/2}}{2\beta^{3/2}}e^{\pi^2/\beta},
\end{equation}
which matches the JT partition function~\cite{Mertens:2022irh} in Eq.~(\ref{eq:Zdisk}) upon the identification $\beta\to\beta/(4\phi_b)$.
This matching also identifies energies to be measured in units of $1/(4\phi_b)$, i.e. $H_{\rm JT}=H_{\rm phys}/(4\phi_b)$.

In summary, the thermal partition function of our pure KS lattice gauge theory with the semigroup measure and the above gravitational boundary prescription reproduces the JT disk partition function.
This is of course no coincidence---the Schwarzian is well known to be related to the Casimir of $\SLR$ in the principal series.
In contrast to previous works, here we are deliberately working on a spatial lattice, i.e. the ``bulk'' states are completely determined by solutions of Gauss's law as discussed above such that the ``holographic'' description in terms of ``boundary'' degrees of freedom follows.
In the next section we will go beyond the limiting case that reproduces JT gravity, and construct generalizations including matter fields based on $\SLR$ that appear most natural from a lattice perspective.

\section{$\SLR$ lattice gauge theory with matter}
\label{sec:matter}

The pure gauge LGT of the previous section is exactly solvable in terms of $\SLR$ degrees of freedom living at the spatial boundary of the 1D lattice.
In this section, we continue along the lines of a conventional KS construction and introduce matter, first as static charges (Sec.~\ref{sec:charges}), then as dynamical degrees of freedom (Sec.~\ref{sec:ws}).
In the static charge case, we again point out connections to JT gravity with matter, which we also employ to benchmark truncated variants of the model in finite-dimensional Hilbert spaces.
In the case of fully dynamical charges, we obtain a genuinely interacting lattice model, and we investigate its dynamics analytically and with numerical simulations.

\subsection{Static charges}
\label{sec:charges}

\subsubsection{Discrete-series charges and a fermionic realization}
\label{sec:discrete}

In order for the charges to couple to links via Gauss's law, they must live in some representation of $\SLR$. 
Specifically, we define charges to live in one of the discrete series of the group, namely in $\mathcal{D}^-_\ell$, with $\ell$ determining the specific irrep.
Note that the choice of $\mathcal{D}^-_\ell$ instead of $\mathcal{D}^+_\ell$ is a convention, and the automorphism $K_1\to-K_1$, $K^\pm\to K^\mp$ of $\slR$ maps $\mathcal{D}^-_\ell$ onto $\mathcal{D}^+_\ell$ and the principal series onto itself with $m\to-m$.
The Casimir of the charges is $\Cas(Q)=\ell(\ell-1)$, and the vertex Hilbert space can be spanned by the basis $\{\ket{\ell,n}\}_{n=0}^\infty$, with the charge operators acting on its elements as 
\begin{align}
        Q&_1\ket{\ell,n}=-(\ell+n)\ket{\ell,n},\nonumber\\
        Q&_{+}\ket{\ell,n}=\sqrt{n(n+2\ell-1)}\ket{\ell,n-1},\\
        Q&_{-}\ket{\ell,n}=\sqrt{(n+1)(n+2\ell)}\ket{\ell,n+1}.\nonumber
\end{align}

Here, $Q_1$ is the elliptic (compact) component. This makes the ladder $n=0,1,...$ a discrete infinite tower that can be understood as a synthetic dimension~\cite{Boada:2011jq} on top of each vertex, as the following fermionic realization shows.

Analogous to the Schwinger representation of angular momentum, we 
consider fermionic mode operators that one can imagine to live in two dimensions: the spatial lattice $x\in{\rm vertices}$ and an additional synthetic $n\geq0$ (see Fig.~\ref{fig:1}{\bf(a)}). Then, the charge operators on a single vertex $x$ 
can be realized as
\begin{equation}
\begin{aligned}
&Q_1=-\sum_{n\geq0}(\ell+n)c_{x,n}^\dagger c_{x,n},\\
&Q_+=\sum_{n\geq0}\sqrt{(n+1)(n+2\ell)}c_{x,n}^\dagger c_{x,n+1},\\
&Q_-=\sum_{n\geq0}\sqrt{(n+1)(n+2\ell)}c_{x,n+1}^\dagger c_{x,n}.
\end{aligned}
\label{eq:fermionic-charges}
\end{equation}

By construction, the one-fermion sector per vertex carries exactly $\mathcal{D}_\ell^-$. 
We also note that a truncation of the tower at $n\le\Nmax$ preserves $[Q_1,Q_\pm]=\pm Q_\pm$ exactly, while
\begin{equation}
  [Q_+,Q_-]=-2Q_{1}-(\Nmax+1)(\Nmax+2\ell)\,c^\dagger_{x,\Nmax}c_{x,\Nmax},
  \label{eq:truncated-algebra}
\end{equation}
so the algebra is broken only through the occupation of the highest level of the tower.
We will later use this truncation in numerical simulations.

\subsubsection{Dressed operators in the presence of static charges}
\label{sec:dressing}
\label{sec:model}
We now impose Gauss's law in the presence of the static charges and solve it exactly. Dressing the charges and the electric fields with the holonomy from the left end of the chain removes the link variables from the Hamiltonian altogether: the model reduces to fermions coupled to a single global flux, with one Casimir of nested fluxes per link. This form is the starting point for both the exact solution of Sec.~\ref{sec:fusion-basis} and the following numerics.

Concretely, the presence of matter in the vertices of the lattice modifies the Gauss-law operators, which now read 
\begin{align}
G_{x,a}=R_{x,a}+L_{x+1,a}+Q_{x,a} \;,
\end{align}
where the label $x$ of the charge operator denotes the vertex between the links $x$ and $x+1$. 
Again, physical states are demanded to satisfy $G_{x,a}\ket{\psi_{\rm phys}}=0$. 
We can solve this equation systematically by invoking the relation between $L_a$ and $R_a$ acting on a group state associated with the link $x$: $R_{x,a}\ket{g_x}=-({\rm Ad}_{g_x})^{b}{}_{a}L_{x,b}\ket{g_x}$. 
The Gauss-law constraint is thus rewritten as $G_{x,a}=-({\rm Ad}_{g_{x}})^{b}{}_{a}L_{x,b}+L_{x+1,a}+Q_{x,a}$. 
Solving iteratively we find 
\begin{equation}
\begin{aligned}
    L_{x,a}=&({\rm Ad}_{g_{1}g_{2}\cdots g_{x-1}})^{b}{}_{a}L_{1,b}\\&-\sum_{m=1}^{x-1}({\rm Ad}_{g_{m+1}g_{m+2}\cdots g_{x-1}})^{b}{}_{a}Q_{m,b}
    \label{eq:Gauss-sol_holonomy}
\end{aligned}
\end{equation}

This relationship can be simplified by defining dressed operators
\begin{equation}
\begin{aligned}
    &\tilde{L}_{x,a}=({\rm Ad}_{P_{x-1}^{-1}})^b{}_aL_{x,b},\\
    &\tilde{Q}_{x,a}=({\rm Ad}_{P_{x}^{-1}})^b{}_aQ_{x,b}.
\end{aligned}
\end{equation}
Here, we defined the holonomy from the left end of the chain up to link $x$ as $P_x=g_1g_2\cdots g_x$, with $P_0=\id$.
Defining $\varepsilon_L=L_1=\tilde{L}_1$, Eq.~\eqref{eq:Gauss-sol_holonomy} becomes
\begin{equation}
    \tilde{L}_{x,a}=\varepsilon_{L,a}-\sum_{m=1}^{x-1}\tilde{Q}_{m,a} \;.
\end{equation}

Given the fermionic realisation of the charge operators in Eq.~\eqref{eq:fermionic-charges}, we can similarly define dressed fermionic operators,
\begin{align}
    \tilde{c}_{x,n}=\sum_{m}D_{nm}^{(\ell)}(P_{x})c_{x,m}.
    \label{eq:dressed-fermions}
\end{align}
These fulfill the anticommutation relations $\{\tilde{c}_{x,n}^\dagger,\;\tilde{c}_{y,m}\}=\delta_{xy}\delta_{nm}$.

One detail remains: since these dressed operators involve the holonomy stemming from the left-most boundary, they do not commute with the left-most electric field $\varepsilon_L$. 
However, we can instead work with the right-most electric field $\varepsilon_{R}=\tilde{L}_{N_{\rm links}}=\varepsilon_L-\sum_{x=1}^{N_{\rm links}-1}\tilde{Q}_x$. Then 
\begin{equation}
    \tilde{L}_x=\varepsilon_R+\sum_{m=x}^{N_{\rm links}-1}\tilde{Q}_m.
    \label{eq:dressed-flux-right}
\end{equation}
The dressing holonomy $P_x$ and the resulting nested fluxes, each given by the right flux plus all the dressed charges to its right, are sketched in Fig.~\ref{fig:1}{\bf(b)}.
Unlike $\varepsilon_L=L_1$, which acts on $g_1$ and hence on every $P_x$, the right-most electric field $\varepsilon_R=(\mathrm{Ad}_{P^{-1}_{N_{\rm links}-1}})^b{}_a L_{N_{\rm links},b}$ is a function of $g_1,\dots,g_{N_{\rm links}-1}$ times the electric field of the last link, which acts only on $g_{N_{\rm links}}$.
The dressed operator $\tilde c_{x,n}$ contains only $c_{x,m}$ and $g_1,\dots,g_x$, with $x\leq N_{\rm links}-1$. Since functions of the group elements commute with each other and with the fermions, the dressed fermionic operators commute with $\varepsilon_R$, and we end up with a theory composed out of fermionic operators and a single independent global $\SLR$ flux. 

\subsubsection{Hamiltonian as a sum of nested Casimirs}
\label{sec:nested}
We have left the Hamiltonian unchanged and now analyse the resulting structure using Eq.~\eqref{eq:dressed-flux-right}. 
First, the Casimirs of dressed and undressed electric fields coincide because 
\begin{equation}
\begin{aligned}
  \Cas(\tilde L_x)&=-\eta^{ab}\tilde L_{x,a}\tilde L_{x,b}\\
  &=-\eta^{ab}({\rm Ad}_{P_{x-1}^{-1}})^c{}_a({\rm Ad}_{P_{x-1}^{-1}})^d{}_b\,L_{x,c}L_{x,d}\\
  &=-\eta^{cd}L_{x,c}L_{x,d}=\Cas(L_x).
\end{aligned}
  \label{eq:casimir-dressed}
\end{equation}
Here we first used that $({\rm Ad}_{P_{x-1}^{-1}})^d{}_b$ depends only on $g_1\cdots g_{x-1}$ and therefore commutes with $L_{x,c}$, which acts on link $x$, and then used that the adjoint action preserves the invariant form, $\eta^{ab}({\rm Ad}_h)^c{}_a({\rm Ad}_h)^d{}_b=\eta^{cd}$ for any $h$, which follows from $\mathrm{tr}(hJ_ah^{-1}\,hJ_bh^{-1})=\mathrm{tr}(J_aJ_b)$. Equation~\eqref{eq:casimir-dressed} is an operator identity.
Therefore, the bare Hamiltonian reads
\begin{equation}
\begin{aligned}
    H_{\rm bare}=\sum_{x=1}^{N_{\rm links}}\Cas(\tilde{L}_x).
\end{aligned}
\end{equation}

Combined with Eq.~\eqref{eq:dressed-flux-right} and using the identity $\Cas(A+B)=\Cas(A)+\Cas(B)+2A\cdot B$ for operators with $[A_a,B_b]=0$, with $A\cdot B=A_1B_1-\half(A_+B_-+A_-B_+)$, the bare Hamiltonian can also be expressed as 
\begin{equation}
\begin{aligned}
    H_{\rm bare}&=N_{\rm links}\Cas(\varepsilon_R)+\sum_{x=1}^{N_{\rm links}-1}x \Cas(\tilde{Q}_x)\\&+2\sum_{x=1}^{N_{\rm links}-1}x\varepsilon_R\cdot\tilde{Q}_x\\&+2\sum_{1\leq x<y\leq N_{\rm links}-1}x\tilde{Q}_x\cdot\tilde{Q}_y.
    \label{eq:dressed-hamiltonian}
\end{aligned}
\end{equation}
The last term, $\tilde{Q}_x\cdot\tilde{Q}_y$, is a long-range interaction between static charges.

\subsubsection{Conserved quantities and block diagonalization}
\label{sec:blocks}

A convenient basis for the physical Hilbert space is $\ket{k_R,m}\ket{n_1,\cdots,n_{N_{\rm links}-1}}$, where $k_R$ and $m$ represent the Casimir label and the elliptic eigenvalue of $\varepsilon_R$, while $n_x$ is the tower level of the dressed fermion at vertex $x$.  The definition of the dressed charges implies that, for $x<y$, the fluxes are related as $\tilde{L}_x=\tilde{L}_y+\sum_{z=x}^{y-1}\tilde{Q}_z$. The extra charges sit at different vertices, so they commute componentwise with $\tilde{L}_y$, and $\Cas(\tilde{L}_y)$ commutes by definition with all components of $\tilde{L}_y$, so that $\left[\Cas(\tilde{L}_x),\Cas(\tilde{L}_y)\right]=0$ and, consequently, $\left[H_{\rm bare}, \Cas(\tilde{L}_x)\right]=0$ $\forall x$. In particular, the $\Cas(\varepsilon_R)$ eigenvalue $-(k_R^2+\frac14)$ is a conserved quantity. By the same argument, the components of $\varepsilon_L$ commute with every $\Cas(\tilde{L}_x)$, so its elliptic eigenvalue, $m-\sum_x(\ell+n_x)$, is conserved as well. Defining $q\equiv m-\sum_x n_x$, the Hamiltonian is block diagonal,

\begin{equation}
  H_{\rm phys}=\int^{\oplus}\!\dd k\,\bigoplus_{q}H_{\rm phys}^{k,q}.
  \label{eq:blocks}
\end{equation}
The evolution generated by $H_{\rm phys}$ hence does not mix different $k_R$ and $q$ sectors, and each of them can be studied independently.

Note that the sector $q$ also determines whether prescription {\it(ii)} of Sec.~\ref{sec:semigroup}, i.e. projecting out states where any link is in a discrete series, acts non-trivially.
The weight of a link with $N_x$ charges to its right is $q+\sum_{y<x}n_y-N_x\ell\geq q-N_x\ell$, so in a sector $q$ all link weights are bounded from below by $q-N_f\ell$, with $N_f$ the number of charges.
Since $\mathcal{P}_k\otimes\mathcal{D}^-_\ell$ contains only principal-series and $\mathcal{D}^-$ channels~\cite{Repka:1978}, and $\mathcal{D}^-$ can only carry negative weights, for $q\geq N_f\ell$ every link remains in the principal series, and the projector of prescription {\it(ii)} is the identity.
Such a condition could not be obtained for situations where combinations of particles in $\mathcal{D}_\ell^+$ and $\mathcal{D}_\ell^-$ are present. 
This is an important distinction to more conventional KS LGTs based on compact groups, where a similar consistency condition does not arise because one can simply work with all irreps that appear in the Peter-Weyl theorem.

\subsubsection{Fusion-chain basis and spectrum}
\label{sec:fusion-basis}
Let us now discuss the integrability of $H_{\rm phys}$ in the presence of static charges and calculate its spectrum. Expressed as a function of Eq.~\eqref{eq:fermionic-charges}, $\Hphys$ in Eq.~\eqref{eq:Hphys} is quartic in fermion operators due to the term proportional to $\tilde{Q}_x\tilde{Q}_y$. While this prevents one from employing free-fermion tools,
the representation theory of $\SLR$ and the nested structure of the Hamiltonian provide enough tools to identify an extensive number of conserved quantities and hence efficiently obtain the exact spectrum.

As shown in Sec.~\ref{sec:blocks}, the nested Casimirs $\Cas(\tilde{L}_j)$ commute with each other and with the Hamiltonian. Since $\Hbare=\sum_j\Cas(\tilde{L}_j)$, its spectrum is fixed by their joint eigenvalues. On physical states living in the semigroup $\SLp$, the links carry only the principal series, as explained in Sec.~\ref{sec:semigroup}, so each $\Cas(\tilde{L}_j)$ takes the value $-(k_j^2+\frac14)$ and the energy of an element of the basis is
\begin{equation}
\begin{aligned}
&\Hphys\ket{k_1,\dots,k_{N_{\rm links}};\alpha}\\&=\frac{1}{N_{\rm links}}\sum_{j=1}^{N_{\rm links}}k_j^2\,\ket{k_1,\dots,k_{N_{\rm links}};\alpha},
\end{aligned}
  \label{eq:fusion-spectrum}
\end{equation}
with $k_{N_{\rm links}}=k_R$ and $\alpha$ the remaining quantum numbers, such as $q$. Since $\tilde{L}_j=\tilde{L}_{j+1}+\tilde Q_j$, the label $k_j$ is obtained by fusing the flux $\Pk{k_{j+1}}$ with the charge $\Dm{\ell}$ at vertex $j$. Thus, it must appear in the decomposition of $\Pk{k_{j+1}}\otimes\Dm{\ell}$. Starting from $k_{N_{\rm links}}=k_R$ and proceeding leftwards, the allowed sequences of labels define a \emph{fusion-chain basis}. The compatible labels $\{k_1,\dots,k_{N_{\rm links}}\}$ form a set of commuting conserved quantities that grows linearly with the system size $N_{\rm links}$.
The change of basis from $\ket{k_R,m}\ket{n_1,\dots,n_{N_{\rm links}-1}}$ is given by the $3j$-symbols of $\SLR$.

\subsubsection{Choice of the Hamiltonian}
\label{sec:H-choice}

We now come back to the discussion at the end of Sec.~\ref{sec:compact} concerning the precise form of the lattice Hamiltonian.
By restricting the sum of Casimir operators to include only the boundary links (with a proper normalisation), we once again connect with the Schwarzian description of JT gravity,
\begin{equation}
    H_{\rm boundary}=-\frac{\Cas(\tilde{L}_1)+\Cas(\varepsilon_R)}{2}-\frac{1}{4}.
    \label{eq:boundary-H}
\end{equation}
Note that this choice leaves the pure gauge partition function unaltered, since in the absence of charges all links carry the same $k$. In the absence of matter, both the bulk-averaged  translation invariant Hamiltonian as well as the explicit boundary Hamiltonian lead to the same physics. 

With matter, however, the bulk-averaged Hamiltonian $H_{\rm phys}$ of Eq.~\eqref{eq:Hphys} and the boundary Hamiltonian $H_{\rm boundary}$ of Eq.~\eqref{eq:boundary-H} differ, since the links between charges carry different labels.
To interpolate between the two cases, we define a one-parameter family of Hamiltonians 
\begin{equation}
\begin{aligned}
    H(g^2)=\left[1-g^2\left(1-\frac{2}{N_{\rm links}}\right)\right]H_{\rm boundary}+\frac{g^2}{N_{\rm links}}H_{\rm int},
    \label{eq:Hg}
    \end{aligned}
    \end{equation}
    where we have defined $H_{\rm int}$ as the sum over the Casimirs of interior links,
    \begin{equation}
    H_{\rm int}=-\sum_{x=2}^{N_{\rm links}-1}\left[\Cas(\tilde{L}_x)+\frac14\right] \;,
    \end{equation}
such that $H(g^2)=(1-g^2)H_{\rm boundary}+g^2 H_{\rm phys}$.
For any $g^2$, \eqref{eq:Hg} commutes with all nested Casimirs, even in the presence of static matter, so that every Hamiltonian in this family is solvable in the fusion-chain basis.

As mentioned above, the limit $g^2=0$ connects to JT gravity, for which the bulk Hamiltonian vanishes on the constraint surface, and the dynamics is generated by boundary terms (Sec.~\ref{sec:jt}).
Additionally, using Eq.~\eqref{eq:dressed-flux-right}, the Gauss laws of all vertices combine into a single global constraint
\begin{equation}
    \varepsilon_{L}-\varepsilon_{R}-\sum_{x=1}^{N_{\rm links}-1}\tilde Q_x=0,
    \label{eq:global-gauss}
\end{equation}
and this is the only impact of the matter charges. 
This structure is reminiscent of the global $\SLR$ constraint that defines the two-sided Hilbert space of JT gravity with matter~\cite{Lin:2019qwu,Lin:2022rbf,Penington:2023dql,Kolchmeyer:2023gqz}, in which the left and right boundary Hamiltonians differ due to the matter between them.
$H_{\rm boundary}$ commutes with every term considered in this work, including the hopping and interaction terms to be added in Sec.~\ref{sec:ws}, when we allow the charges to move dynamically.

For $g^2=1$ we recover the bulk-averaged Hamiltonian, for which the interior links carry electric energy, which has no counterpart in JT gravity. 
In this case, a hopping term no longer commutes with the Hamiltonian due to the interior nested Casimirs entering its definition, thus constituting a qualitative distinction between the $g^2=0$ and $g^2>0$ cases. 
Since the latter case appears more natural from a conventional lattice perspective, one might speculate about reasons that could select the JT limit dynamically.

\subsubsection{Truncations}
\label{sec:schemes}

\begin{figure*}
    \centering
    \includegraphics[width=0.85\textwidth]{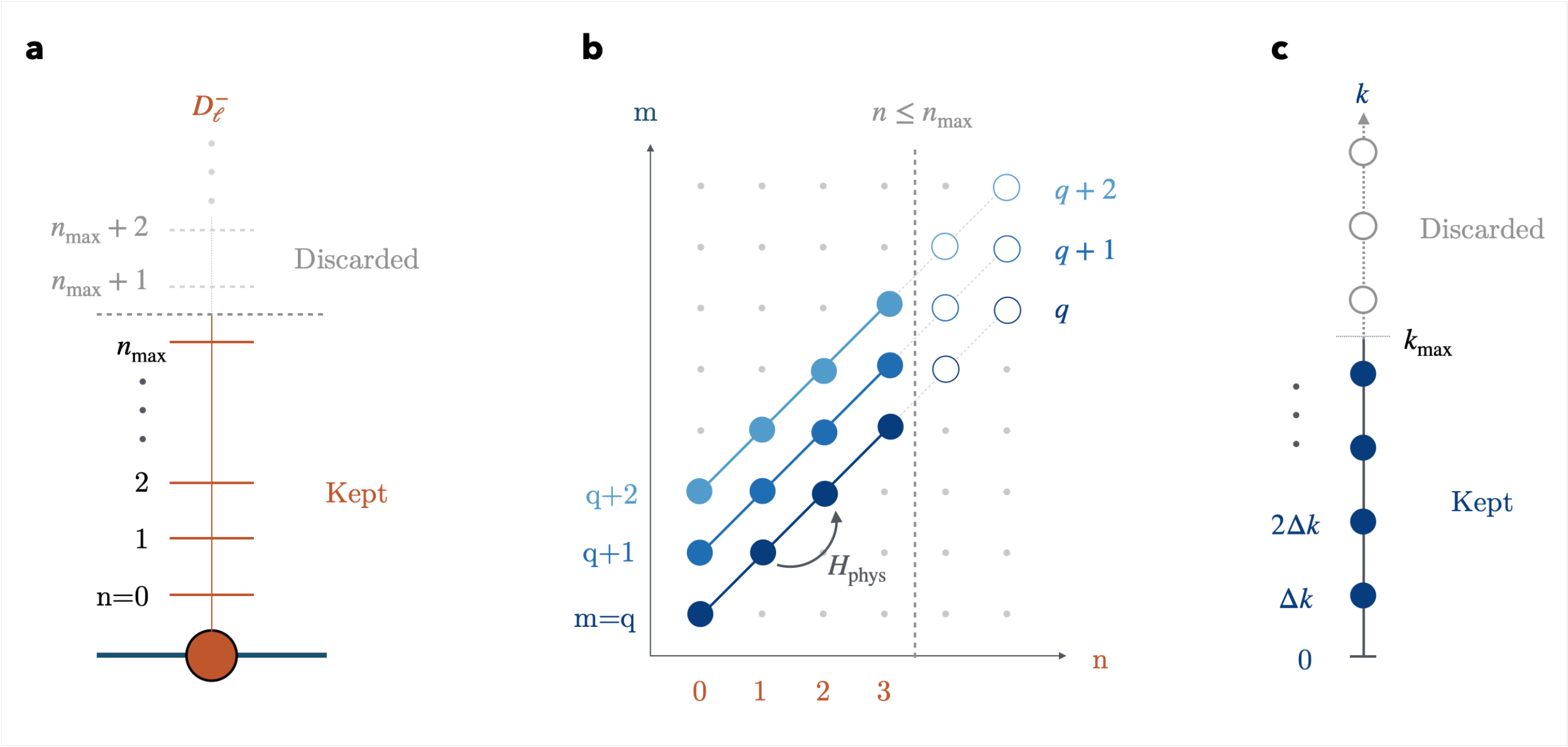}
    \caption{{\bf(a)}~Towers of the charges cut at $n\le\Nmax$. {\bf(b)}~Flux index $m=q+\sum_xn_x$ left free. {\bf(c)}~Grid $k_i=i\,\Delta k$ in the continuous label.}
    \label{fig:2}
\end{figure*}

To study the Hamiltonian $H_{\rm phys}$ (and later interacting generalizations) numerically, we employ two truncation schemes: {\it (i)} impose a maximum value $n_{\rm max}$ on the towers $\{n_x\}_{x=1}^{N_{\rm links}-1}$ such that $n_x\leq n_{\rm max}$ $\forall x$ (see Fig.~\ref{fig:2}{\bf(a)}); and {\it (ii)} discretize the continuum $k_R$ on a grid $k_i=i\Delta k$ and impose a maximum value $k_{\rm max}$ (see Fig.~\ref{fig:2}{\bf(c)}). Since the Hamiltonian is block-diagonal, the second regularization just determines the values of $k_R$ that one can study separately in order to compute observables for which $k_R$ must be integrated. 
The former truncation, in contrast, has effects within every $H_{k,q}$. We expect to recover the untruncated results in the limit $n_{\rm max}, k_{\rm max}\to\infty$ and $\Delta k\to0$. 
To test this, we study the convergence of selected observables as $n_{\rm max}, k_{\rm max}$ are increased and $\Delta k$ is decreased in order to determine them up to a required precision. 

Additionally, since the Hamiltonian is block-diagonal in $q=m-\sum_x n_x$ as well, we can
study it sector by sector: within a sector, $m$ is fixed by the tower levels (see Fig.~\ref{fig:2}{\bf(b)}), so the only
free labels are the $\{n_x\}$, and each $H^{k,q}_{\rm phys}$ has dimension
$(n_{\rm max}+1)^{N_{\rm links}-1}$. 
While the overall space grows exponentially with system size, the nested structure shown in Sec.~\ref{sec:fusion-basis} allows one
to bypass this growth.

\subsubsection{Benchmark: density of states and matter vertex of JT gravity}
\label{sec:bench-fusion}

Next we benchmark our lattice construction against JT gravity in the presence of matter for two links, where $H_{\rm int}=0$ such that $H(g^2)$ becomes independent of $g^2$ and equal to $H_{\rm boundary}$.
We select the static charge in the only vertex as a fixed irrep $\mathcal{D}^-_\ell$ and prepare its internal state in $n=0$. 
On the other hand, we fix the label $k_R$ of the flux $\varepsilon_R$, and work also in a fixed sector $q=m-n$, both of which are conserved as outlined in Sec.~\ref{sec:blocks}. For $n=0$, the elliptic eigenvalue of $\varepsilon_R$ is $m=q$. Then, the remaining degree of freedom is the label $k_L$ of the left link, whose flux is $\tilde{L}_1=\varepsilon_R+\tilde{Q}_1$. We obtain it by constructing the left-link Casimir operator $\Cas(\tilde{L}_1)=\Cas(\varepsilon_R+\tilde{Q}_1)=-(k_R^2+1/4)+\ell(\ell-1)+2\varepsilon_R\cdot\tilde{Q}_1$ and diagonalising it. For $q\geq\ell$, the case considered here, all its eigenvalues are of the form $-(k_L^2+1/4)$, i.e. the fusion contains only principal-series channels; for $q<\ell$, discrete-series channels also appear, which are absent in the Plancherel decomposition of $\SLp$ and which we discuss at the end of this section. It is from these eigenvalues that we extract the set of $k_L$ and study its distribution. The initial state is not an eigenstate of $\Cas(\tilde{L}_1)$, and its distribution in $k_L$ can be connected to JT observables
that are known in closed form.
As we show below, it contains the density of states of the left region and the matter vertex, the two ingredients of the JT two-point function~\cite{Mertens:2022irh,Blommaert:2018iqz}. In our numerics, we also directly assess the effect of truncating to a finite $\Nmax$.

Since the eigenvalues of $\Cas(\tilde{L}_1)$ correspond to $-(k_L^2+1/4)$, we can define the operator $T=-\Cas(\tilde{L}_1)-\frac14$, whose eigenvalues are directly $k_L^2$. For two links, $T$ is an infinite tridiagonal matrix in the basis $\ket{n}\equiv\ket{k_R,m=q+n}\ket{\ell,n}$, with diagonal and off-diagonal elements $\braket{n|T|n}=a_n$ and $\braket{n+1|T|n}=b_n$ given, respectively, by
\begin{equation}
\begin{aligned}
    a_n&=2(n+q)(n+\ell)+k_R^2-\ell(\ell-1),\\
    b_n&=\sqrt{\left[\left(n+q+\half\right)^2+k_R^2\right](n+1)(n+2\ell)}.
\end{aligned}
\label{eq:jacobi}
\end{equation}

The eigenvectors $\ket{k_L}$ of $T$, $T\ket{k_L}=k_L^2\ket{k_L}$, have components $v_n=\braket{n|k_L}$ which follow a recurrence equation given by the tridiagonal structure of $T$, $b_{n-1}v_{n-1}+a_nv_n+b_nv_{n+1}=k_L^2v_n$, with $v_{-1}=0$. Solving for $v_{n+1}$, this gives
\begin{equation}
    v_{n+1}=\frac{(k_L^2-a_n)v_n-b_{n-1}v_{n-1}}{b_n}.
    \label{eq:tridiagonal_recurrence}
\end{equation}
Solving this recurrence equation gives any component $n$ of a given eigenvector as a function of its $0$-th component. Specifically, since each iteration of Eq.~\eqref{eq:tridiagonal_recurrence} multiplies by $k_L^2-a_n$ once, a component $n$ is given by a degree-$n$ polynomial of $k_L^2$ times $v_0$, $v_n=p_n(k_L^2)\,v_0$. The specific polynomials are determined by the form of $a_n$ and $b_n$ and, comparing Eq.~\eqref{eq:jacobi} with the recurrences tabulated in Ref.~\cite{Koekoek:2010}, they are found to be continuous dual Hahn polynomials. The only remaining undetermined parameter is $v_0$. It can be fixed by noting that both the $\ket{n}$ and the $\ket{k_L}$ bases resolve the identity, $\sum_n\ket{n}\bra{n}=\int\dd k_L\ket{k_L}\bra{k_L}$. Taking the $nm$ component of this matrix identity, we find $\delta_{nm}=\int\dd k_L\braket{n|k_L}\braket{k_L|m}=\int\dd k_L\,p_n(k_L^2)\,p_m(k_L^2)\,|\braket{0|k_L}|^2$, such that $v_0$ is given, up to a phase, by the weight that makes the continuous dual Hahn polynomials orthonormal. 
This weight is unique and known analytically~\cite{Koekoek:2010}. It takes the form
\begin{equation}
\begin{aligned}
  w_q(k_L;k_R)\equiv|v_0|^2&=\frac{\rho(k_L)}{\pi^2}\,\gamma_\ell^2(k_L,k_R)\,R_q(k_L,k_R),
\end{aligned}
\label{eq:wq}
\end{equation}
normalised as $\int_0^\infty\dd k_L\,w_q=1$.

Three ingredients enter this expression: First, we recognize $\rho(k_L) =k\sinh2\pi k$ as the JT density of states of the left region. Second,
\begin{align}
\gamma_\ell^2&=\frac{\Gamma(\ell\pm ik_L\pm ik_R)}{\Gamma(2\ell)}
\end{align}
is the matter vertex of JT gravity~\cite{Mertens:2022irh,Blommaert:2018oro} (the $\pm$ notation is a shorthand for the product over the four combinations of signs). 
And third, 
\begin{align}
    R_q&=\frac{|\Gamma(q+\frac12-\ell+ik_L)|^2}{|\Gamma(q+\frac12+ik_R)|^2}
\end{align}
depends only on the state of the right flux, which in our case is the elliptic eigenstate $m=q$.
In JT gravity, the same role would be played by the parabolic boundary state of Sec.~\ref{sec:bench-Z}, for which the amplitude is exactly $\rho\,\gamma_\ell^2$~\cite{Blommaert:2018oro}. 

\begin{figure}
  \centering
  \includegraphics[width=0.92\columnwidth]{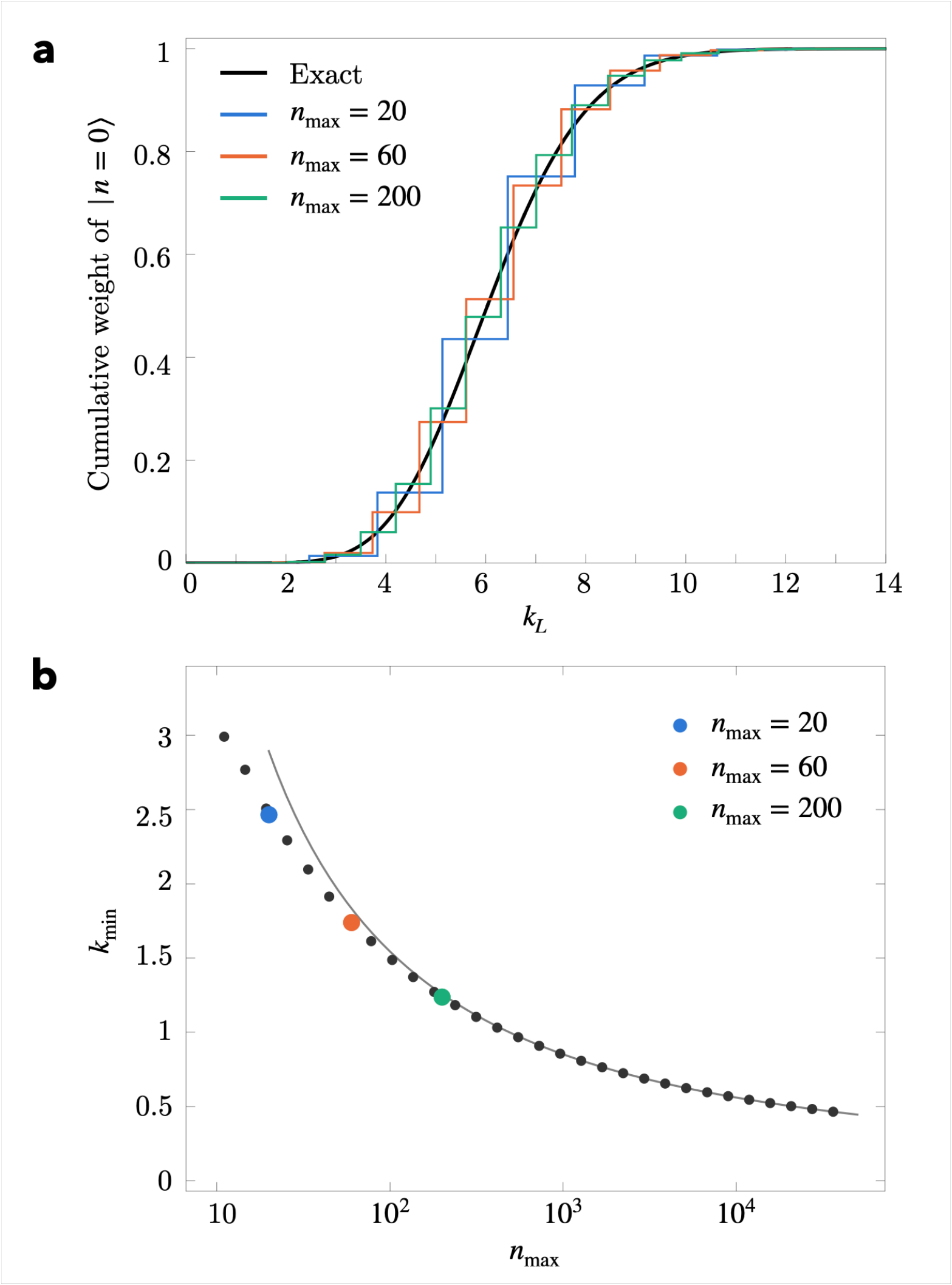}
  \caption{Distribution of the left label $k_L$ in the state $|n=0\rangle$, for $\ell=5$, $q=6$ and $k_R=0$. {\bf(a)}~Cumulative distribution: exact, from the numerical integral of Eq.~\eqref{eq:wq} (black), and truncated block with $\Nmax=20$, $60$ and $200$ (staircases, one step per eigenvalue). {\bf(b)}~Lowest resolved momentum $k_{\min}$, the square root of the lowest eigenvalue of the truncated block, against $\Nmax$; coloured points: the cutoffs of panel (a). Grey line: analytical prediction, Eq.~\eqref{eq:kmin-cond}, valid up to corrections of relative order $1/\Nmax$. Eq.~\eqref{eq:kmin-cond} is derived in Appendix~\ref{app:kmin}.}
    \label{fig:3}
\end{figure}

In Fig.~\ref{fig:3}(a) we plot as a solid line the integral $\int^{k_L}_0\dd\kappa\,w_q(\kappa;k_R)$, and compare it to the cumulative sum of the numerical weights $|v_0|^2$ obtained after exactly diagonalizing $T$ truncated at $n\leq n_{\rm max}$, for $n_{\rm max}=20,60,200$. The exact curve crosses every step, since the truncated block is the Gauss quadrature rule of $w_q$~\cite{Golub:1969}, and the effect of truncation is to limit the resolution of allowed $k_L$ values: the higher $n_{\rm max}$, the denser the resulting grid of $k_L$. The lowest resolved value, $k_{\min}$, decreases only logarithmically with $\Nmax$, asymptotically as $k_{\min}\simeq\pi/\ln(\Nmax/n_*)$ with $n_*$ fixed by $(q,\ell,k_R)$ (see Appendix~\ref{app:kmin} for details) and compare it with exact diagonalisation in Fig.~\ref{fig:3}(b). 
We conclude that $\Nmax$ must grow exponentially in $1/k_{\min}$ to resolve small values of $k_L$.

For $q<\ell$, $w_q$ also contains isolated points at negative values of $k_L^2$~\cite{Koekoek:2010}, which are the discrete-series channels of the fusion $\mathcal{P}_{k_R}\otimes\mathcal{D}^-_\ell$ in $\SLR$. These channels are absent from the Plancherel decomposition of $\SLp$, so in cases where they are present, we should project them out before any further computations. 

\subsection{Dynamical matter}
\label{sec:ws}

So far, inspired by JT gravity and the structure that the inclusion of matter generates, we have considered an LGT where matter is introduced only as static charges. The effect of these charges is to act as sources in Gauss's law, such that regions of links separated by a static charge have different values of $k$, the principal series label. Since the matter does not affect the form of the Hamiltonian, which remains a sum of nested, mutually commuting Casimirs, the system has an extensive number of conserved quantities and can thus be solved efficiently. 
We now 
abandon the JT interpretation and add terms that have no established gravitational counterpart, while retaining the non-compact structure whose consequences we set out to study.

\subsubsection{Interacting model}
One of the simplest terms that we can add to the Hamiltonian is a gauge-invariant hopping term,
\begin{equation}
    H_{\rm hop}=-\thop\sum_{x=1}^{N_{\rm links}-2}c_{x,n}^\dagger D_{nm}^{(\ell)}(g_{x+1})c_{x+1,m}+{\rm h.c.},
\end{equation}
where $D_{nm}^{(\ell)}(g_{x+1})$ are the matrix elements, in $\mathcal{D}^-_\ell$, of the group element $g_{x+1}$ of link $x+1$, which connects the vertices $x$ and $x+1$, and a sum over $n,m$ is implied.

In our 1D lattice model it is convenient to work with the dressed fermions of 
Eq.~\eqref{eq:dressed-fermions} rather than with the bare ones. Inserting them into the hopping Hamiltonian, we find
\begin{equation}
\begin{aligned}
    H_{\rm hop}&=-\thop\sum_{x,a,b}\tilde{c}^\dagger_{x,a}D_{an}^{(\ell)}(P_{x})D_{nm}^{(\ell)}(g_{x+1})\\
    &\qquad\times D_{mb}^{(\ell)}(P_{x+1}^{-1})\tilde{c}_{x+1,b}+{\rm h.c.}\\
    &=-\thop\sum_{x,n}\tilde{c}^\dagger_{x,n}\tilde{c}_{x+1,n}+{\rm h.c.},
\end{aligned}
    \label{eq:hop}
\end{equation}
since $P_{x+1}=P_xg_{x+1}$. The hopping term is therefore completely independent of the holonomy, so the model still contains only fermions living in one physical and one synthetic dimension, plus a global flux living in the principal series of $\SLR$.

Local self-interactions of the matter are also straightforward to include,
\begin{align}
H_{\rm U}&= U\sum_{x}\left(\sum_n\tilde{c}^\dagger_{x,n}\tilde{c}_{x,n}\right)^2 = U\sum_{x}\left(\sum_n {c}^\dagger_{x,n}{c}_{x,n}\right)^2 \;.
\end{align}
Note that, being completely local, this interaction is naturally gauge-invariant and takes the same form with bare or dressed fermions.

As a central result of this work, we arrive at the following model
\begin{equation}
    H=H(g^2)+H_{\rm hop} + H_U \;,
\end{equation}
representing a manifestly gauge-invariant, interacting $\SLR$ lattice gauge theory.
In particular, the hopping removes one key property of $H_{\rm phys}$: since $\left[\Cas(\tilde{L}_i),\,\Cas(\tilde{L}_j)\right]=0$ for all $i,j$, we had $\left[\Cas(\tilde{L}_i),\,H_{\rm phys}\right]=0$ for all $i$, which allowed us to diagonalize the Hamiltonian using representation theory as outlined in Sec.~\ref{sec:fusion-basis}. The inclusion of $H_{\rm hop}$ spoils this strategy, since $\left[\Cas(\tilde{L}_i),\,H_{\rm hop}\right]\neq0$ for $i$ belonging to the bulk of the chain. In other words, a fermion hopping across link $i$ changes $\tilde{L}_i$ as usual in KS LGTs. Therefore, the inclusion of $H_{\rm hop}$ makes the system no longer exactly solvable for generic $g^2>0$, and only the $g^2=0$ case, corresponding to JT (for $U=0$), remains integrable because the boundary Casimirs commute with the hopping. 

\subsubsection{Effective description at large $U$}
In contrast to the 
case with static charges, the chain is now divided into domains of constant $k$, separated by moving fermions.
We leave a more detailed study of this model for future work, and for now focus on the case of large $U\rightarrow \infty$.
In this case
one site is occupied by at most one fermion, which allows us to derive a simpler  
effective Hamiltonian. 
To this end, note that 
the total number of fermions $N_f =  \sum_{x,n} c_{x,n}^\dagger c_{x,n}$ is a conserved quantity and corresponds to $N_f+1$ regions
on the lattice, each classified by a constant value of $k$. 
Since the Casimirs of the boundary links commute with the full Hamiltonian, we can fix their values to $k_L$ and $k_R$. 
We thus obtain $N_f+1$ string-like regions, labelled by conserved fluxes $k_1=k_L, k_2, \dots, k_{N_f+1}=k_R$, 
but their physical size remains dynamical.

\begin{figure}
    \centering
    \includegraphics[width=0.92\linewidth,height=0.7\textheight,keepaspectratio]{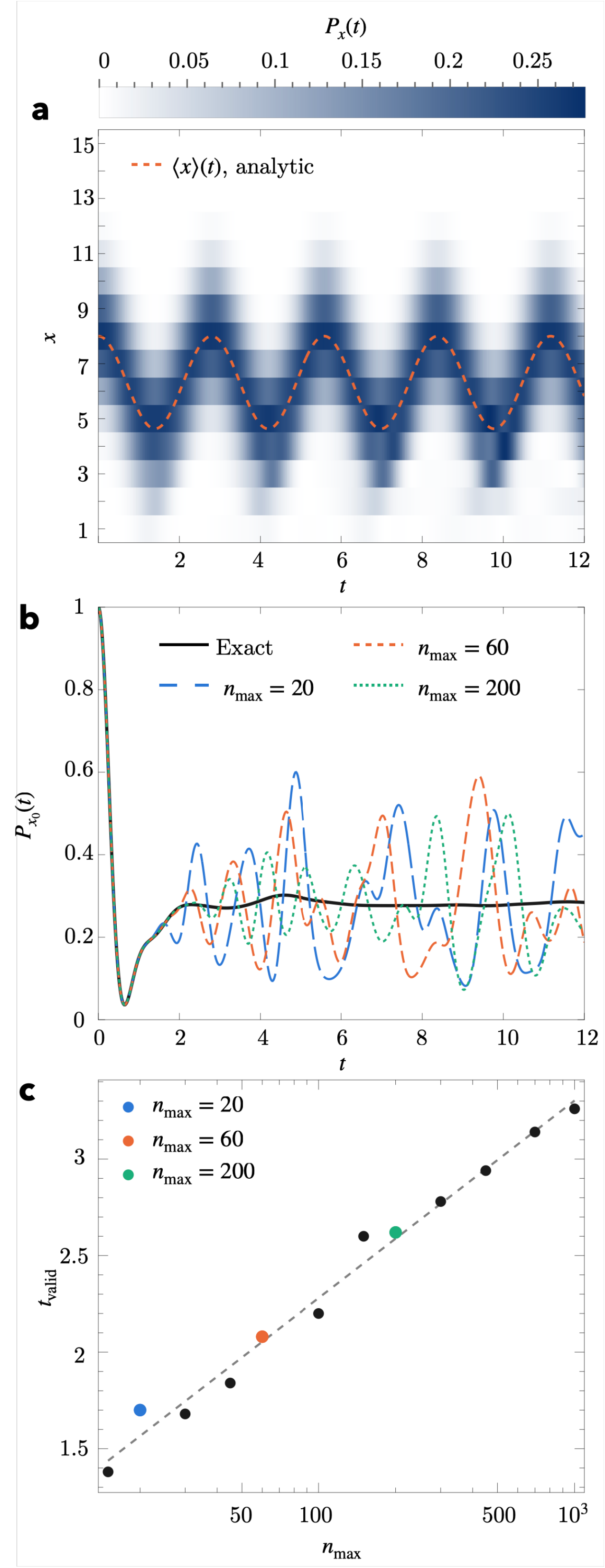}
    \caption{One mobile charge ($q=6$, $\ell=5$, $k_R=0$, $N_{\rm links}=16$, $g^2=1$).
\textbf{(a)}~Bloch oscillations at fixed labels $(k_L,k_R)=(6,0)$ of a Gaussian packet
centred at $x_0=8$; dashed: analytic $\langle x\rangle(t)$.
\textbf{(b)}~Return probability of $|x_0\rangle\otimes|n=0\rangle$: exact average over
$k_L$ (solid) versus the truncated model (dashed).
\textbf{(c)}~Time up to which the truncated model agrees with the exact curve within
$10^{-2}$, with logarithmic fit $t_{\rm valid}\simeq0.44\ln n_{\max}+0.24$.}
    \label{fig:4}
\end{figure}

At large $U$, the interaction $H_U$ dynamically enforces the absence of double occupancies, i.e. locally $\sum_n c_{x,n}^ \dagger c_{x,n} \le 1$.
Consequently the fermions never change their relative ordering on the chain, so the corresponding string regions remain well-defined under unitary time evolution, and the flux label $k_a$ of the region between fermions $a-1$ and $a$ is fixed by the Casimir of $\varepsilon_R+\sum_{b\geq a}\tilde{Q}^{(b)}$, with $\tilde{Q}^{(b)}$ the dressed charge of the $b$-th fermion from the left. Since the hopping~\eqref{eq:hop} is diagonal in the tower index and preserves the order of the fermions, it also commutes with these Casimirs, which, unlike $\Cas(\tilde{L}_i)$ for a fixed link $i$, do not depend on the positions of the fermions. Assuming that the position of the $a$-th fermion is given by $x_a$, with $x_0=0$ and $x_{N_f+1}=N_{\rm links}$, the corresponding electric energy is given by 
\begin{equation}
\begin{aligned}
    E(x_1,\dots,x_{N_f})&=\frac{1}{N_{\rm links}}\sum_{a=1}^{N_f+1}(x_a-x_{a-1})k_a^2\\
    &=\sum_{a=1}^{N_f}F_a x_a+k_R^2,
\end{aligned}
\end{equation}
with a ``force''
\begin{equation}
    F_a=\frac{k_a^2-k_{a+1}^2}{N_{\rm links}}.
    \label{eq:force}
\end{equation}
This is the diagonal part of the Hamiltonian in the basis of positions, to which $H_{\rm hop}$ adds an off-diagonal part. Dropping the constant $k_R^2$ (and the contribution $UN_f$ of $H_U$), we obtain 
\begin{equation}
\begin{aligned}
    H_{\rm eff}&=\sum_{a=1}^{N_f}F_{a}x_a\\&-\thop\sum_a\left(\ket{\dots,x_a+1,\dots}\bra{\dots,x_a,\dots}+{\rm h.c.}\right),
    \end{aligned}
\end{equation}
with an implicit hard-core constraint that no two particles occupy the same site. 
Here we impose this restriction dynamically via large $U$ 
in order to keep the model classically tractable.
For finite $U$, nothing prevents two fermions with different $n$ 
to occupy the same vertex, in which case the labels are no longer conserved (Sec.~\ref{sec:double}). 
Note also that $H_{\rm eff}$ involves no truncation, since the tower index enters only through the conserved labels.

The system therefore consists of a set of fermions subject to linear potentials due to connecting flux strings, with a strength given by a two-sided ``tug-of-war'' between the principal-series labels $k_a^2$ of neighbouring regions. In other words, we obtain a set of dynamically coupled Wannier--Stark ladders~\cite{Glueck:2002vyi}. In the language of Sec.~\ref{sec:H-choice}, Eq.~\eqref{eq:force} corresponds to the bulk-averaged Hamiltonian, $g^2=1$; for general $g^2$, the forces are $F_a=(g^2/N_{\rm links})(k_a^2-k_{a+1}^2)$, so that they vanish at the JT point $g^2=0$, where the charges travel freely. At fixed labels and fixed $g^2$, the forces also vanish for long chains as $1/N_{\rm links}$. This is indeed the expected behaviour if we interpret the chain as a lattice discretisation at fixed physical length $L$ with lattice spacing $a=L/N_{\rm links}$, such that the force per unit length, $F_a/a=g^2(k_a^2-k_{a+1}^2)/L$, is independent of $N_{\rm links}$.

\subsubsection{One fermion}

Let us consider a single fermion, $N_f = 1$.
The system then consists of only two string regions, labelled by $k_L$ and $k_R$. 
In the following, we work in the $q=m-n$ sector and fix the right flux $k_R$.
The remaining gauge-invariant degree of freedom can either be encoded in the left link state $k_L$ or in the tower state of the charge, $n$. 

First, let us fix the left flux $k_L$, such that the charge is in a superposition of all the tower states $n$. 
Depending on the balance between the fixed $k_R$ and $k_L$, the charge is then subjected to a left- or a right-oriented force, set by $F\equiv F_1=(k_L^2-k_R^2)/N_{\rm links}$. 

To illustrate this, we take a chain of $N_{\rm links}=16$ links with $\thop=2$, $k_L=6$ and $k_R=0$, and consider the evolution of an initial Gaussian packet around the vertex $x_0=8$ with standard deviation $\sigma_0=1.5$ sites.
The resulting probability $P_x(t)$ to find the charge on site $x$ is shown in Fig.~\ref{fig:4}{\bf(a)}, as a function of evolution time $t$.
We observe a clear Bloch oscillation with period $T_B=2\pi/|F|$, with
mean position accurately described by the analytical prediction
\begin{equation}
    \braket{x}(t)=x_0-\frac{2\thop}{F}e^{-1/(8\sigma_0^2)}(1-\cos Ft).
    \label{eq:bloch-pos}
\end{equation}

For a charge initially localised on a single vertex, $\sigma_0\to0$, the probability is instead given by 
\begin{equation}
    P_x(t)=\mathcal{J}_{x-x_0}^2\left(\frac{4\thop}{F}\sin\frac{Ft}{2}\right) \;.
    \label{eq:bloch-prob}
\end{equation}
In this case the expectation value of $x$ stays pinned to its initial value, $\braket{x}(t)=x_0$, and the effect of the force $F$ is a Bloch breathing: the probability $P_x(t)$ is symmetric around $x_0$ but broadens and shrinks periodically in time, with frequency $F$.

\begin{figure}[!tbp]
    \centering
    \includegraphics[width=0.92\linewidth,height=0.7\textheight,keepaspectratio]{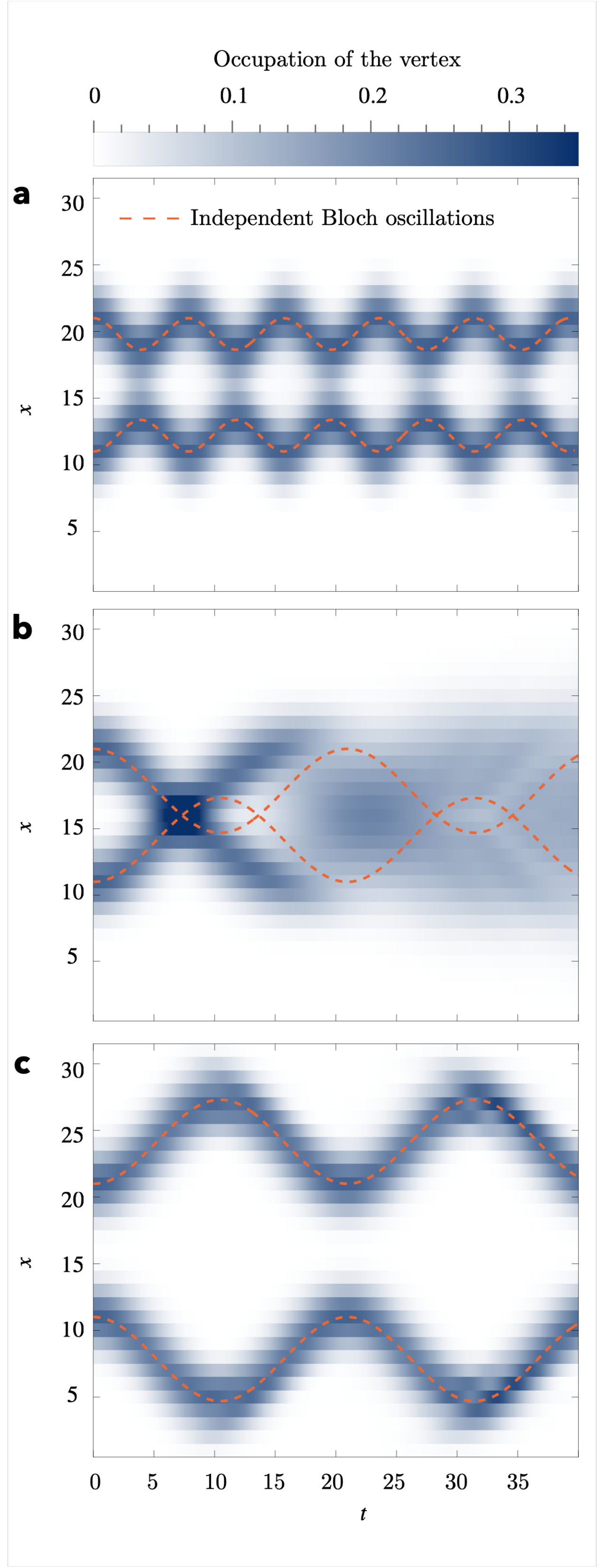}
    \caption{Two mobile charges, initially in Gaussian packets of width $\sigma_0=1.5$ centred at $x=11$ and $x=21$, with $k_L=k_R=4$, $N_{\rm links}=32$ and $\thop=0.5$, for $g^2=1$. Colour: occupation of each vertex; dashed: independent Bloch oscillations of each charge. \textbf{(a)}~$k_{\rm mid}=6.45$ ($F=0.8$). Since $4J_K/F\simeq5<r_0$, the charges turn back before meeting. \textbf{(b)}~$k_{\rm mid}=5.06$ ($F=0.3$). Since $4J_K/F\simeq13>r_0$, the charges collide and form a mobile bound pair. Note that the dashed curves are independent Bloch oscillations, not the exact two-body result. \textbf{(c)}~$k_{\rm mid}=2.53$ ($F=-0.3$). They move apart and oscillate independently.}
    \label{fig:5}
\end{figure}

Next, we consider the case where $k_L$ is not fixed, but instead we constrain the state of the charge to be $\ket{n=0}$. 
The left links are therefore in a superposition of $k_L$ values, with the weight given by $w_q(k_L;k_R)$ in Eq.~\eqref{eq:wq}. Initially, the fermion is at position $x=x_0$. Then each $k_L$ value describes an independent channel with force $F=(k_L^2-k_R^2)/N_{\rm links}$, which vanishes for $k_L=k_R$ and is right- or left-oriented for $k_L<k_R$ or $k_L>k_R$, respectively. Since the channels are conserved and the initial state is a superposition of them, the probabilities of the channels add, weighted by $w_q$; in the untruncated theory (i.e. $\Nmax\to\infty$) there is a continuum of $k_L$ values, so this sum becomes an integral. 
For example, we can compute the return probability of the particle to $x=x_0$, given by 
\begin{equation}
    P_{x_0}(t)=\int_0^\infty \dd k_L\,w_q(k_L;k_R)\,\big|\braket{x_0|e^{-iH_{\rm eff}(k_L)t}|x_0}\big|^2.
\end{equation}

Since every channel is equivalent to a Wannier--Stark model with a different force, each will generate Bloch oscillations with a different frequency. After integrating over the channels, these oscillations dephase, and the return probability relaxes to a plateau ($\simeq0.28$). This is illustrated by the black solid curve shown in 
Fig.~\ref{fig:4}{\bf(b)}.
If we instead impose a truncation $n\leq\Nmax$, the allowed $k_L$ will not define a continuum but instead 
take the $\Nmax+1$ discrete values given by the eigenvalues of the truncated matrix $T$, as explained in Sec.~\ref{sec:bench-fusion}, and their weights are the squared first components of the corresponding eigenvectors. 
The integral is thus approximated by its Gauss quadrature, resulting in the dashed lines in Fig.~\ref{fig:4}{\bf(b)}.
The effect of this truncation 
clearly shows that 
agreement with the exact result holds only up to a time $t_{\rm valid}$, which we estimate here  as the first time at which the two curves differ by more than $10^{-2}$. 
After that, spurious revivals appear.
This time increases with $\Nmax$ as expected, and we find that the dependence is logarithmic, $t_{\rm valid}\simeq0.44\ln\Nmax+0.24$ (see Fig.~\ref{fig:4}{\bf(c)}).
This scaling is consistent with the spacing $\sim\pi/\ln\Nmax$ of the discrete $k_L$ values (Sec.~\ref{sec:bench-fusion} and Appendix~\ref{app:kmin}).

\subsubsection{Two fermions at $U\rightarrow \infty$}

The same analysis can be generalised to an arbitrary number of fermions $N_f\leq N_{\rm links}-1$: in the absence of double occupation ($U \rightarrow \infty$), there are $N_f+1$ regions of constant principal series label $k_a$. 
Without this restriction, the model  is substantially more complex since the coincidence of multiple fermions in the same site would invoke recoupling coefficients of $\SLR$ ($6j$-symbols) which can change the value of $k_a$ for intermediate regions, invalidating the construction of $H_{\rm eff}$. 

Here, we study the phenomenology that arises in the absence of double occupation. 
Additionally, we restrict our attention to two fermions, $N_f = 2$. The case of larger $N_f$ follows analogously.
As for the case of a single fermion, there are two natural types of initial states.

First, we could fix the right-most flux principal label $k_R$ 
plus the tower state of all $N_f$ internal charges, e.g. $n_x=0$ for all $x$. 
This state would not be an eigenstate, but a superposition of channels, and, as for one fermion, each charge dephases. 
Alternatively, we may fix all of the links' principal labels in a fusion-chain fashion, so that the dynamics takes place in a single channel. 
We now study the latter scenario and distinguish three regimes with different balances between fluxes, as shown in Fig.~\ref{fig:5}.
For all cases 
we set $k_L=k_R=4$, so that the value of the inner flux $k_{\rm mid}$ determines their balance: for $k_{\rm mid}<4$, the particles will be initially repelled, whereas for $k_{\rm mid}>4$ there will be an initial attraction. Following Eq.~\eqref{eq:force}, $F_2=-F_1\equiv F=(k_{\rm mid}^2-k_L^2)/N_{\rm links}$, so $H_{\rm eff}=F(x_2-x_1)$ plus hopping: $F>0$ is thus attractive and $F<0$ is repulsive.

We find that this two-body problem has an exact solution.
For $k_L=k_R$, the electric energy is $F(x_2-x_1)$, up to a constant, and the hard-core constraint can be written as $x_2-x_1\geq1$.
One can thus introduce a centre-of-mass coordinate, $X=\frac12(x_1+x_2)$, and a relative coordinate, $r=x_2-x_1$, in terms of which the electric energy simply becomes $Fr$. 
Since $X$ does not affect the energy, we can define a quasi-momentum $K$, conjugate to $X$, which is conserved with periodic boundary conditions or for infinite chains, or far from the boundaries of the chain. 
The relative coordinate then obeys a Wannier--Stark equation with an effective hopping $J_K=2t_{\rm h}\cos(K/2)$, together with a hard-wall boundary condition at $r=0$. 
This leads to 
\begin{equation}
-J_K(\psi_{r+1}+\psi_{r-1})+Fr\psi_r=E\psi_r,\qquad\psi_0=0 \;,
\label{eq:relative}
\end{equation}
which can be solved using the recurrence relation of the Bessel functions. Normalisability at large $r$ selects $\psi_r\propto\mathcal{J}_{r-\nu}(2J_K/F)$, with the energy given by $E=F\nu$. 
The wall at $r=0$ quantises $\nu$, and thus also $E$, through the constraint
\begin{equation}
\mathcal{J}_{-\nu}(4t_{\rm h}\cos(K/2)/F)=0.
\label{eq:quantisation}
\end{equation}

For $F=0$, Eq.~\eqref{eq:relative} has a continuous band $\left[-2|J_K|,2|J_K|\right]$, so that the charges propagate freely.
For any $F\neq0$, however, the spectrum is discrete and every eigenstate is bound. 
The $n$-th state is localised around $r\approx\nu_n$, with width $\sim2|J_K|/|F|$. Without the wall, i.e.\ for $r\in\mathbb{Z}$, normalisability at $r\to-\infty$ would force $\nu_n=n$ exactly. 
This is the conventional Wannier--Stark ladder, with $K$-independent energies and both charges performing independent Bloch oscillations. 
This situation is approximately realised for states localised far from the wall, $r\approx\nu_n\gg2|J_K|/|F|$. 
Such a state has negligible amplitude at $r=0$, so the condition $\psi_0=0$ is approximately satisfied. These states therefore satisfy $\nu_n\approx n$ and are well described as independent Bloch oscillations. 
In contrast, states with $\nu_n\lesssim2|J_K|/|F|$ overlap with the $r=0$ wall, and their $\nu_n$ deviate from the integers by a $K$-dependent amount. 
Their energies $E_n(K)=F\nu_n(K)$ therefore disperse, so these states describe mobile bound pairs whose centre of mass propagates. 

Which regime is realised depends on the initial state. 
In Fig.~\ref{fig:5}, the initial state consists of two Gaussian packets of width $\sigma_0=1.5$ centred at $x_1=11$ and $x_2=21$, such that $r_0=10$, $K\approx0$ and $J_K\simeq2t_{\rm h}$. 
Since the packets start at rest, the relative motion starts at the bottom of the band, with kinetic energy $\simeq-2J_K$, so that $E\simeq Fr_0-2J_K$. Since $E=F\nu$, the initial state populates mainly the states with $\nu\simeq r_0-2J_K/F$, i.e.\ centred one width downhill from $r_0$. 
For $F<0$, as in Fig.~\ref{fig:5}(c), these lie further from the wall than $r_0$, and the charges oscillate independently. 
For $F>0$ they are far from the wall only if $r_0-2J_K/F\gg2J_K/F$, i.e.\ $r_0\gg4J_K/F\simeq8t_{\rm h}/F$. 
This ratio is $5<r_0$ in Fig.~\ref{fig:5}(a), where the charges oscillate independently, and $13>r_0$ in Fig.~\ref{fig:5}(b), where they form a mobile bound pair. 

We stress that this solution holds only for the hard-core effective Hamiltonian with $k_L=k_R$. 
In the full model, $k_{\rm mid}$ is not conserved during the collision and no such solution is available.
We leave a thorough investigation of the full interacting model for future work. 

\section{Conclusions}
\label{sec:conclusions}

\subsection{Discussion}

In this work we have carried out a bottom-up construction of a Hamiltonian lattice gauge theory (LGT) in one spatial dimension based on the non-compact group $\SLR$.
We have paid careful attention to the features that distinguish it from the more conventional compact case of Kogut-Susskind (KS) LGTs. 
In particular, the following observations may serve as a blueprint for more general constructions involving non-compact groups.

First, we have set up the gauge field Hilbert space in the elliptic basis, where the infinite-dimensional unitary representations create an infinite ladder of states on which the generators act as nearest-neighbour hopping.
This is a necessary workaround because in a hyperbolic basis the algebra cannot be carried even approximately by a discretisation. 
Similar obstructions are to be expected for non-compact directions in more general groups.

Second, we introduced charges in the discrete series $\mathcal{D}^-_\ell$, which can be realised by fermions in an infinite synthetic dimension. More general non-compact groups will again require dealing with irreps that are necessarily infinite-dimensional.
In our simple 1D example, the standard trick to ``integrate out'' all but a background gauge field degree of freedom again goes through: dressing the fermions with the holonomy from the boundary, we effectively removed the link variables.
Consequently, the model reduces to fermions coupled to a single global flux, with a Hamiltonian given by a sum of nested, mutually commuting Casimir operators. 
Similar to the more familiar case of non-Abelian compact groups, the spectrum then follows from a fusion-chain basis, with an extensive set of conserved labels.

Our 1D LGT has a clear physical anchor.
Restricting the links to the positive semigroup $\SLp$, we showed how to directly reproduce two exact results known from the quantization of Jackiw--Teitelboim gravity. 
Namely, the disk partition function, through the Plancherel measure $k\sinh 2\pi k$, and, from the fusion of a flux with a charge, the product of the density of states and the matter vertex. 
With this benchmark in mind, we showed that the truncation of the charge tower at $n_{\max}$ turns this distribution into its Gauss quadrature, whose lowest resolved momentum decreases only logarithmically, $k_{\min}\simeq\pi/\ln(n_{\max}/n_*)$.
We attribute the relatively slow convergence of this ``naive'' truncation (when compared to the more conventional compact case) to the non-compact nature of $\SLR$. 
In contrast, for compact gauge groups and bosonic modes truncations are well under control: a cutoff growing only polylogarithmically suffices~\cite{Jordan:2012xnu,Tong:2021rfv}.
Our finding thus suggests that the Hilbert space discretization of non-compact groups poses additional practical problems for both classical and quantum simulations.

Finally, departing from an established gravitational interpretation, we added further gauge-invariant hopping and interaction terms.
At strong coupling, in the absence of double occupations, we find that flux string labels of the regions between charges remain conserved, and the model maps onto coupled Wannier--Stark ladders whose forces are set by the balance of neighbouring fluxes. 
A single charge thus displays Bloch oscillations, and averaging over the continuum of channels, these oscillations dephase.
Here a hard Hilbert space cutoff limits the time of validity of the simulation to very short times, $t_{\rm valid}\propto\ln n_{\max}$. 
For two charges, 
we solve the strong coupling limit exactly, which identifies a condition for particles to oscillate independently.
Otherwise both particles form a bound state that cannot be accounted for with independent Bloch oscillations. 
Lifting the hard-core constraint 
would not preserve the interior principal series labels, 
and we anticipate much richer non-integrable dynamics.

\subsection{Outlook}
\label{sec:double}

In our constructed 1D LGT, arguably the most interesting scenario is that of Fig.~\ref{fig:5}{\bf(b)}, where fermions are close together, 
but in the case of finite $U$, such that the simple 
effective Hamiltonian is not valid anymore.
This drastically increases the difficulty of the problem, since then the Casimir operators of the interior regions would not commute with $H_{\rm phys}+H_{\rm hop}$, thus reducing the number of conserved charges and increasing the overall complexity. 
We plan to revisit this scenario in a future work.

In a gravitational interpretation, our lattice model should give access to boundary correlators of JT gravity, such as the two-point function and its interpretation in terms of the wormhole length.
It would be interesting to couple to additional degrees of freedom that would enable the corresponding black hole to evaporate.
By leveraging the lessons learned in quantum simulation of compact (non)-Abelian LGTs~\cite{Wiese:2013uua,Banuls:2019bmf,Bauer:2022hpo,DiMeglio:2023nsa,Halimeh:2023lid}, our approach thus paves a new path toward quantum simulating JT gravity via a non-compact Kogut-Susskind LGT. 
In this context, the logarithmic scaling of simulation time and resulting exponential scaling of resources in the hard cutoff regularization poses particularly strong restrictions and motivates the study of alternative discretizations of non-compact groups.

A natural follow-up, also with respect to possible quantum simulations of black-hole evaporation, is the construction of Hamiltonian LGTs with the centrally extended Poincaré group $\widehat{ISO}(1,1)$ instead of $\SLR$ as a noncompact gauge group, as motivated by the type of dilaton theory introduced by Callan, Giddings, Harvey and Strominger (CGHS)~\cite{Callan:1992rs,Cangemi:1992bj}.
In this context, a close relation to the complex SYK model~\cite{Afshar:2019axx} could offer benchmarks analogous to the ones of JT gravity discussed in the present article.

More generally, our work lays the basis for developing Hamiltonian LGTs based on non-compact gauge groups in higher spatial dimensions.
Specifically, our gauge-invariant constructions of principal series gauge fields coupled to discrete series matter fields generalizes straightforwardly.
The main remaining step is the construction of magnetic plaquette interactions that would give rise to propagating pure gauge field degrees of freedom in higher dimensions.

Finally, we note that also continuum Yang-Mills theory is typically restricted to compact groups.
Given the relatively simple topological interpretation of JT gravity, we have only studied bare lattice models in the present work and have not attempted to obtain a continuum limit directly from the lattice.
In the present 1D example, one might speculate whether or not the topological JT limit is the only sensible continuum limit that emerges dynamically.
To answer this question one should study the full phase diagram, which in turn raises the question what phases to expect for a non-compact gauge group in the first place, both in one as well as in higher spatial dimensions.
Our exploratory study for $\SLR$ can also serve as a starting point for numerical tensor network simulations in this direction.

 \begin{acknowledgments}
The authors thank R.~Ott, D.~Gonz\'alez-Cuadra, A.~Bermudez and J.~Beltr\'an Jim\'enez for useful discussions.
C.F.-C.\ acknowledges the hospitality of IQOQI and the University of Innsbruck during his research stay (April--June 2026), supported by the Ayuda Econ\'omica para Personal Investigador Predoctoral 2025 of the Programa Becas Santander (Universidad Aut\'onoma de Madrid).
C.F.-C.\ acknowledges support from the European Union's Horizon Europe research and innovation programme under grant agreement No.~101114305 (MILLENION-SGA1 EU Project), from PID2024-161474NB-I00 (MCIU/AEI/FEDER, UE), and from QUITEMAD-CM TEC-2024/COM-84.
C.F.-C. also acknowledges support from the Grant IFT Centro de Excelencia Severo Ochoa CEX2020-001007-S funded by MCIN/AEI/10.13039/501100011033, and from the CSIC Research Platform on Quantum Technologies PTI-001. This publication has been funded within the framework of the R\&D\&I Project CEX2025-001574-S, funded by MICIU/AEI/10.13039/501100011033. The research presented in this publication falls within the research line Quantum.
T.V.Z.\ is supported by an ERC Starting grant (QS-Gauge, 101220401). Funded by the European Union. Views and opinions expressed are however those of the author(s) only and do not necessarily reflect those of the European Union or the European Research Council Executive Agency. Neither the European Union nor the granting authority can be held responsible for them.
The authors acknowledge the use of generative AI tools (Anthropic, Claude Opus models accessed via Claude Code, 2026; and OpenAI, ChatGPT 6, 2026) in the preparation of this work, in particular to cross-check analytical and numerical derivations, to survey the literature and verify bibliographic references, and to improve the clarity of the manuscript’s wording. All AI-assisted derivations were independently verified by the authors analytically and numerically. All AI-assisted text, equations, figures, code, and references were reviewed and, where necessary, revised by the authors, who take full responsibility for the content of this article.
\end{acknowledgments}

\appendix
\section{Unitary irreducible representations of $\SLR$}
\label{sec:irreps}

The irreducible representations (irreps) can be classified according to the Casimir eigenvalue, while the diagonalised generator's eigenvalues further classify the states within it. The two remaining generators define ladder operators whose action is fixed by the algebra (up to phases). From Eqs.~\eqref{eq:K-algebra} and~\eqref{eq:casimir}, $K^\mp K^\pm=K_1(K_1\pm1)-\Cas$, and since $(K^\pm)^\dagger=K^\mp$ in a unitary representation, we get
\begin{equation}
  \big\|K^\pm\ket{m}\big\|^2=m(m\pm1)-\Cas\;\ge 0 .
  \label{eq:unitarity}
\end{equation}
Different unitary representations satisfy this inequality in different ways.
We need the following two families for this work.

\emph{Principal series} $\Pk{k}$, with $\lambda=\half+ik$ and $k\ge0$; the labels $k$ and $-k$ give the same Casimir and equivalent representations~\cite{Iliesiu:2019xuh}.
In this case,
$\Cas=\lambda(\lambda-1)=-(k^2+\tfrac14)$, and the right-hand side of Eq.~\eqref{eq:unitarity} is $(m\pm\half)^2+k^2>0$ for $m\in\mathbb{Z}$, so the spectrum is unbounded. We have
\begin{equation}
\begin{gathered}
  K_1\ket{k,m}=m\ket{k,m},\\   K^\pm\ket{k,m}=\sqrt{\big(m\pm\tfrac12\big)^2+k^2}\;\ket{k,m\pm1}.
\end{gathered}
\label{eq:principal}
\end{equation}

\emph{Discrete series} $\mathcal{D}^\pm_{\lambda=\ell}$, $\ell=1,2,\dots$.
For
$\Cas=\ell(\ell-1)$ the right-hand side of Eq.~\eqref{eq:unitarity} factorises as $(m\pm\ell)(m\mp\ell\pm1)$ and vanishes at $m=\mp\ell$, bounding the spectrum of the diagonalised generator.
The states in the irrep $\Dm{\ell}$ can be written as $\ket{\ell,n}$, with $n=0,1,2,\dots$ and $m=-(\ell+n)$, yielding
\begin{equation}
\begin{aligned}
  K^+\ket{\ell,n}&=\sqrt{n\,(n+2\ell-1)}\;\ket{\ell,n-1},\\   K^-\ket{\ell,n}&=\sqrt{(n+1)(n+2\ell)}\;\ket{\ell,n+1}.
\end{aligned}
\label{eq:discrete}
\end{equation}
$\mathcal{D}^+_\ell$ is its mirror image with lowest weight $+\ell$.

There are two further classes of unitary reps, the so-called complementary series and the trivial one~\cite{Bargmann:1947,Lang:1985}.
As these do not appear in the Plancherel decomposition of square-integrable functions on the group, they play no role in this article.

\section{JT gravity, BF theory and their constraints}

\label{sec:JT}

\subsection{JT gravity}
Jackiw--Teitelboim (JT) gravity is a two-dimensional theory of gravity coupled to a scalar field commonly known as the \emph{dilaton}. A general dilaton theory has an action of the form~\cite{2d-dilaton}
\begin{equation}
    S=\int_{\mathcal{M}}\dd^2x\, \sqrt{-g}\left(\phi R-V(\phi)\right),
\end{equation}
in units $16\pi G_N=1$, where $\phi$ is the dilaton field, $g$ the determinant of the metric and $R$ its Ricci scalar. The integration is over a spacetime manifold here denoted as $\mathcal{M}$. The function $V(\phi)$ determines the specific dilaton theory, with JT corresponding to $V(\phi)=-2\phi$, which forces the metric to be anti-de Sitter (AdS). This action can be translated into the Hamiltonian formulation by expressing the metric in the ADM parametrisation~\cite{ADM}
\begin{equation}
    \dd s^2=e^{2\omega}\left[-\sigma^2\dd t^2+(\dd x+M\dd t)^2\right],
    \label{eq:adm-metric}
\end{equation}
with fields $\omega(t,x)$, $\sigma(t,x)$ and $M(t,x)$. The final Hamiltonian can be written as~\cite{LGK}
\begin{equation}
    H=\int \dd x\,(M\mathcal{F}+\sigma\mathcal{G}),
    \label{eq:ham-pure-dilaton}
\end{equation}
where we abbreviated 
\begin{equation}
\begin{aligned}
\mathcal{F} &= \omega'\Pi_\omega+\phi'\Pi_\phi-\Pi_\omega',\\
\mathcal{G} &= 2\phi''-2\phi'\omega'-\half\Pi_\phi\Pi_\omega+e^{2\omega}V(\phi) \;.
\end{aligned}
\label{eq:JT-constraints}
\end{equation}
Here a prime denotes a spatial derivative and $\Pi_\phi$ and $\Pi_\omega$ are the momenta conjugate to $\phi$ and $\omega$.
In the language of quantization of constrained systems, $\mathcal{F}$ and $\mathcal{G}$ are first-class constraints, and $H$ vanishes on the constraint surface.
Since first-class constraints generate gauge transformations, one can interpret $\mathcal{F}$ and $\mathcal{G}$ as the \emph{momentum} and \emph{Hamiltonian} constraints, generating spatial diffeomorphisms and time translations, respectively.

\subsection{BF theory}
The classical JT theory is known to be equivalent to a $\slR$ BF theory~\cite{Fukuyama:1985gg,Isler:1989hq,Chamseddine:1989yz}.
The latter has the action 
\begin{equation}
    S_{BF}=\int \dd^2x\,{\rm tr}(\varphi F_{tx})=\half\int \dd^2x\,\eta_{ab}\varphi^aF_{tx}^b.
\end{equation}
Here $F^a_{tx}=\dot A_x^a-(A_t^a)'+f^a_{bc}A_t^bA_x^c$, with $f^a_{bc}$ the structure constants of the $\slR$ algebra \eqref{eq:real-algebra}, is a component of the field strength tensor associated to a gauge field $A_\mu=A_\mu^a J_a$, and $\varphi=\varphi^a J_a$ is a scalar field in the adjoint representation. The connection to a first-order formulation of JT gravity follows from identifying $A_\mu^0=\Omega_\mu$, where $\Omega_\mu$ is the spin connection of JT, and $A_\mu^a=e_\mu^a$ for $a=1,2$, where $e_\mu^a$ is the zweibein used in the first-order formalism, connected to the metric components by the relation $g_{\mu\nu}=-e^1_\mu e^1_\nu+e^2_\mu e^2_\nu$. We also fix the local Lorentz freedom by choosing $e^1_x=0$. For the ADM metric Eq.~\eqref{eq:adm-metric} this leads to $e^1_t=\sigma e^\omega$, $e^1_x=0$, $e^2_t=Me^\omega$ and $e^2_x=e^\omega$.
Analysing the BF action from the perspective of its constraints, we find the three components of Gauss's law as
\begin{equation}
    G_a=\eta_{ab}(\varphi^b)'+\eta_{bc}\varphi^bf^c_{ad}A_x^d \;,
\end{equation}
which must vanish on the constraint surface. The $0$th component, $G_0=(\varphi^0)'-e^\omega\varphi^1$, and the condition $e^1_x=0$ constitute a pair of second-class constraints that can be solved explicitly as $\varphi^1=\frac{(\varphi^0)'}{e^\omega}$. Then, upon identification of $\varphi^0=4\phi$, $\varphi^2=\frac{2\Pi_\omega}{e^\omega}$ and $\Omega_x=-\Pi_\phi/2$, the components $G_1$ and $G_2$ become
\begin{equation}
    \begin{aligned}
        &G_1=-2e^{-\omega}\mathcal{G},\\
        &G_2=-2e^{-\omega}\mathcal{F},
    \end{aligned}
\end{equation}
matching the JT constraints \eqref{eq:JT-constraints}.
In summary, after fixing the local Lorentz freedom, the remaining BF Gauss constraints reproduce the momentum and Hamiltonian constraints of JT gravity.

\subsection{Including matter}
The inclusion of matter in JT gravity leads to the same structure as in Eq.~\eqref{eq:ham-pure-dilaton}
but with modified constraints $\mathcal{F}\to\mathcal{F}+\mathcal{P}_{\rm matter}$ and $\mathcal{G}\to\mathcal{G}+\mathcal{E}_{\rm matter}$, where $\mathcal{P}_{\rm matter}$ and $\mathcal{E}_{\rm matter}$ are the momentum density and energy density of the matter fields. Therefore, the inclusion of matter only modifies the first-class constraints, and hence Gauss's law of the BF theory, but it does not induce any extra non-zero term to the Hamiltonian.
This motivates the construction with only static charges in the main text, where matter is introduced on the vertices, Gauss's law is modified but no further term is added to $H_{\rm phys}$. 

\section{Lowest resolved momentum of the truncated tower}
\label{app:kmin}

In this Appendix, we compute the lowest resolved momentum $k_{\min}$ of the truncated block of Sec.~\ref{sec:bench-fusion}, i.e.\ the minimum value of $k_L$, as a function of $n_{\rm max}$. This is equivalent to studying $n_{\rm max}$ as an IR regulator that modifies the size of the box in which $k_L$ is quantised. This dependence can be predicted analytically in the large-$n$ limit, where, to leading order, $a_n\simeq2n^2$ and $b_n\simeq n^2$, while the subleading terms combine into $a_n-b_n-b_{n-1}=-\frac14+\mathcal{O}(1/n)$. Now defining $v_n=(-1)^nf(n)$, the recurrence of the eigenvector components can be written as
\begin{equation}
\begin{aligned}
    k_L^2f(n)={}&(a_n-b_n-b_{n-1})f(n)\\
    &-b_n[f(n+1)-f(n)]\\
    &+b_{n-1}[f(n)-f(n-1)],
\end{aligned}
\end{equation}
which, if we treat $n$ as continuous and denote $x=k_L^2$, yields the differential equation 
\begin{equation}
xf=-\frac{\dd}{\dd n}\left(n^2\frac{\dd f}{\dd n}\right)-\frac14f.
\end{equation}

Upon the changes of variable $u=\ln n$ and $f=e^{-u/2}g$, we end up with $g''+xg=0$, where the prime denotes differentiation with respect to $u$, with solution $g=A\cos(k_L u+\varphi)$, with $\varphi$ a phase, i.e. $v_n\simeq(-1)^nA\,n^{-1/2}\cos(k_L\ln n+\varphi)$. From normalisation, the amplitude is $A=\sqrt{\frac{2}{\pi}}$. The phase $\varphi$ is not fixed by the differential equation, since it depends on the region of small $n$, where the continuum approximation fails. It is fixed instead by the exact solution. With $\alpha=q+\frac12-\ell$ and $\beta=\ell+ik_R$, the polynomials $p_n$ have the closed form~\cite{Koekoek:2010}
\begin{equation}
\begin{aligned}
  p_n(k_L^2)={}&(-1)^n\sqrt{\frac{|(\alpha+\beta)_n|^2}{n!\,(2\ell)_n}}\\
  &\times{}_3F_2\!\left(\begin{matrix}-n,\ \alpha+ik_L,\ \alpha-ik_L\\ \alpha+\beta,\ \alpha+\bar\beta\end{matrix};1\right),
\end{aligned}
\label{eq:pn-closed}
\end{equation}
with $(x)_n=\Gamma(x+n)/\Gamma(x)$ the Pochhammer symbol. At large $n$, the terminating ${}_3F_2$ is dominated by two terms~\cite{Long:2020}, proportional to $n^{-\alpha\mp ik_L}$, which are complex conjugates of each other and therefore add up to a single cosine. Multiplying by $v_0=\sqrt{w_q}$, all constants cancel, and we recover the cosine solution, including the amplitude $A=\sqrt{2/\pi}$, with $\varphi(k)=\arg c(k)$ and
\begin{equation}
  c(k)=\frac{\Gamma(2ik)}{\Gamma(\alpha+ik)\,\Gamma(\beta+ik)\,\Gamma(\bar\beta+ik)}.
  \label{eq:phase}
\end{equation}

The truncation at $n=n_{\rm max}$ is implemented by the condition $p_{n_{\rm max}+1}(k_L^2)=0$. Inserting the cosine solution, this holds for $k_L\ln n_{\rm max}+\arg c(k_L)=\frac{\pi}{2}+j\pi$, with $j=0,1,\dots$, and the lowest eigenvalue, $j=0$, gives 
\begin{equation}
k_{\rm min}\ln n_{\rm max}+\arg c(k_{\min})=\frac{\pi}{2},
\label{eq:kmin-cond}
\end{equation}
so $n_{\rm max}=\exp\left(\frac{\pi/2-\arg c(k_{\rm min})}{k_{\rm min}}\right)$. In Fig.~\ref{fig:3}(b), we present solutions to this equation, plotted as a function of $n_{\rm max}$, and compare it with the numerical $k_{\rm min}$ obtained after numerically diagonalizing $T$. As expected, the large-$n$ prediction is more exact as $n_{\rm max}$ increases, with deviations of relative order $1/n_{\rm max}$ ($17\%$ at $n_{\rm max}=20$, $0.55\%$ at $n_{\rm max}=640$), but it correctly describes the dependence also for small $n_{\rm max}$. Using $\Gamma(z)\simeq\frac{1}{z}-\gamma_E$ near $z=0$, with $\gamma_E$ the Euler constant, and $\ln \Gamma(x+ik)\simeq\ln\Gamma(x)+ik\psi(x)$, with $\psi(x)$ the digamma function, we end up with $\arg c(k) \simeq-\frac{\pi}{2}-k\ln n_{*}$ at small $k$, with $\ln n_*=2\gamma_E+\psi(q+\half-\ell)+2{\rm Re}\,\psi(\ell+ik_R)$ ($n_*\simeq67$ for the parameters of Fig.~\ref{fig:3}). Then, $\cos(k\ln n+\arg c(k))\simeq \sin\big(k\ln(n/n_*)\big)$. The condition of vanishing $p_{n_{\rm max}+1}$ turns into $k_{\rm min}\ln\frac{n_{\rm max}}{n_*}=\pi$, so that $k_{\rm min}\simeq\frac{\pi}{\ln(n_{\rm max}/n_*)}$, or equivalently $n_{\rm max}\simeq n_*\exp(\pi/k_{\rm min})$. This form is approached only slowly, since the neglected terms are of relative order $k_{\rm min}^2$, but it captures the essential point: $n_{\rm max}$ must grow exponentially in $1/k_{\rm min}$ to resolve smaller values of $k_L$, which agrees with the behaviour shown in Fig.~\ref{fig:3}(b).

\bibliography{references}

\end{document}